\documentclass[usenatbib]{mnras}
\usepackage{lineno}
\usepackage{natbib}
\usepackage{graphicx}
\usepackage{amssymb}
\usepackage{color}
\usepackage{upgreek}
\usepackage{amsmath}
\usepackage{gensymb}

\DeclareMathOperator{\sech}{sech}
\newcommand{\ptrsa}{RSPTA}

\newcommand{\sa}{SvA}

\author[Perera et al.]{B.~B.~P.~Perera,$^{1}$\thanks{E-mail: bhakthiperera@gmail.com} E.~D.~Barr,$^{2}$ M.~B.~Mickaliger,$^{1}$ A.~G.~Lyne,$^{1}$ D.~R.~Lorimer,$^{3,4}$
\newauthor B.~W.~Stappers,$^{1}$ R.~P.~Eatough,$^{2}$ M.~Kramer,$^{2,1}$ C.~Ng,$^{5}$ R.~Spiewak,$^{6}$ M.~Bailes,$^{6,7}$
\newauthor D.~J.~Champion,$^{2}$ V.~Morello$^1$ and A.~Possenti$^8$\\
\\ $^1$ Jodrell Bank Centre for Astrophysics, School of Physics and Astronomy, The University of Manchester, Manchester M13 9PL, UK
\\ $^2$ Max-Planck-Institut f$\ddot u$r Radioastronomie, Auf dem H$\ddot u$gel 69, D-53121 Bonn, Germany
\\ $^3$ Department of Physics and Astronomy, West Virginia University, Morgantown, WV 26506, USA
\\ $^4$ Center for Gravitational Waves and Cosmology, West Virginia University, Chestnut Ridge Research Building, Morgantown, \\WV 26505, USA
\\ $^5$ Department of Physics and Astronomy, University of British Columbia, 6224 Agricultural Road, Vancouver, BC V6T 1Z1, Canada
\\ $^6$ Centre for Astrophysics and Supercomputing, Swinburne University of Technology, PO Box 218, Hawthorn, VIC 3122, Australia
\\ $^7$ ARC Centre of Excellence for Gravitational Wave Discovery (OzGrav), Swinburne University of Technology, PO Box 218, \\VIC 3122, Australia
\\ $^8$ INAF - Osservatorio Astronomico di Cagliari, via della Scienza 5, I-09047 Selargius (CA), Italy}

\title[Dynamics of Galactic centre pulsars]{The dynamics of Galactic centre pulsars: constraining pulsar distances and intrinsic spin-down}

\date{}

\begin{document}

\maketitle

\begin{abstract}
Through high-precision radio timing observations, we show that five recycled pulsars in the direction of the Galactic Centre (GC) have anomalous spin period time derivative ($\dot P$) measurements -- PSRs~J1748$-$3009, J1753$-$2819, J1757$-$2745, and J1804$-$2858 show negative values of $\dot P$ and PSR~J1801$-$3210 is found to have an exceptionally small value of $\dot P$. We attribute these observed $\dot P$ measurements to acceleration of these pulsars along their lines-of-sight (LOSs) due to the Galactic gravitational field. Using models of the Galactic mass distribution and pulsar velocities, we constrain the distances to these pulsars, placing them on the far-side of the Galaxy,  providing the first accurate distance measurements to pulsars located in this region and allowing us to consider the electron density along these LOSs. We find the new electron density model YMW16 to be more consistent with these observations than the previous model NE2001. The LOS dynamics further constrain the model-dependent intrinsic $\dot P$ values for these pulsars and they are consistent with measurements for other known pulsars. In the future, the independent distance measurements to these and other pulsars near the GC would allow us to constrain the Galactic gravitational potential more accurately. 
\end{abstract}


\section{Introduction}
\label{intro}

Pulsars are among the most reliable macroscopic periodic sources in the universe and their long-term stability is comparable to that of atomic clocks \citep[e.g.][]{pt96,hcm+12}. This allows the parameters of pulsars to be measured to exquisite precision through measuring and modelling the arrival times of their pulses \citep[e.g.][]{ktr94}. These measurements are sensitive to any dynamical changes caused by external forces on the pulsar such as the gravitational forces exerted by other masses \citep[e.g.][]{frk+17,psl+17a}.

Pulsars are thought to be powered by the loss of rotational kinetic energy and thus, gradually spin down over time, leading to positive values of $\dot P$. The current pulsar population strongly supports this hypothesis with measured positive $\dot P$ values for the vast majority of pulsars\footnote{\url{http://www.atnf.csiro.au/people/pulsar/psrcat}} \cite[see][]{mht+05}. However, some pulsars, located in globular clusters, show significant negative $\dot P$ measurements \citep[e.g][]{wkm+89,frk+17,jcj+06,cpl+06,lfrj12}. These $\dot P$ values are not entirely intrinsic, but induced by the dynamics due to the LOS component of the acceleration of the pulsar, mainly in the gravitational potential of the cluster itself, locating them on the far-side of the cluster \citep[see, e.g.,][]{fck+03,frk+17}. 

Similar to the globular clusters, the Galaxy has a strong gravitational potential within the central regions due to the high stellar and gas densities. We therefore expect negative $\dot P$ measurements from pulsars that are located near the Galactic centre (GC) on the far-side of the Galaxy (see \S~\ref{los_acc} for details). However, prior to this work, there was  no conclusive evidence for this. As we show below using our timing observations, PSRs~J1748$-$3009 \citep{kek+13}, J1753$-$2819 \citep[originally announced by][]{mic13}, J1757$-$2745 \citep{ncb+15}, and J1804$-$2858 \citep[][Barr et al., {\it in prep}]{mbc+19} all show significant negative $\dot P$ measurements, and PSR~J1801$-$3210 \citep{nbb+14} shows a remarkably small but positive $\dot P$ measurement when compared to the known pulsar population.

In addition to the pulsars listed above, we note that our timing of PSR~J1746$-$2758 \citep{ncb+15} does not show an unusual $\dot P$ measurement, however, its location in the Galaxy and some measured properties have similarities to those of above mentioned pulsars, making it useful to consider in this work. The previously reported rotational, and some other measured and derived parameters, for these pulsars are given in Table~\ref{psr_info}, noting that four of them had no previous $\dot P$ measurements. All six pulsars considered here are located along LOSs which pass near the GC with an angular separation of $<5^\circ$. With their comparably large dispersion measure (DM -- which accounts for the frequency-dependent time delay of the radio pulses due to electrons in the inter-stellar medium along the LOS) values, it is likely that these pulsars are located within the inner regions of the Galaxy around the GC \citep{cl02,ymw17}. Given the special locations of these pulsars, it is likely that their unusual observed $\dot P$ measurements are not entirely intrinsic, but rather induced by the pulsar acceleration in the Galactic gravitational potential. We further note that these pulsars are useful in studying the DM environment of the inner region of the Galaxy.

\begin{table*}
\begin{center}
\caption{
The previously published parameters of the pulsars in this study. The distance is estimated using the electron density models given in \citet{ymw17}(YMW16) and \citet{cl02}(NE2001). Since there are no multi-frequency observations in this study, we use these published DM measurements and keep them fixed in our timing models.     
 }
\label{psr_info}
\begin{tabular}{lrrrccccc}
\hline
\multicolumn{1}{c}{PSR} &
\multicolumn{1}{c}{$l$} &
\multicolumn{1}{c}{$b$} &
\multicolumn{1}{c}{$P$} &
\multicolumn{1}{c}{$\dot{\rm P}$} &
\multicolumn{1}{c}{DM} &
\multicolumn{1}{c}{Ref.} &
\multicolumn{1}{c}{$D_{\rm YMW16}^{\dagger\dagger}$} &
\multicolumn{1}{c}{$D_{\rm NE2001}^{\dagger\dagger}$} \\
\multicolumn{1}{c}{ } &
\multicolumn{1}{c}{($^\circ$)} &
\multicolumn{1}{c}{($^\circ$)} &
\multicolumn{1}{c}{(ms)} &
\multicolumn{1}{c}{(s/s)} &
\multicolumn{1}{c}{(cm$^{-3}$~pc)} &
\multicolumn{1}{c}{ } &
\multicolumn{1}{c}{(kpc)} &
\multicolumn{1}{c}{(kpc)} \\
\hline
\hline
J1746$-$2758$^\dagger$ & 0.97 & 0.49 & 487.53 & -- & 422.0 & 1 & 4.2&5.2 \\
J1748$-$3009 & 359.27 & $-$1.15 & 9.68 & -- & 420.2 & 2 & 5.1&5.1 \\
J1753$-$2819 &    1.46    & --1.25  & 18.62     & -- &    298.0  & 3  &   4.1&4.5   \\
J1757$-$2745$^\dagger$ & 2.40 & $-$1.72 & 17.69 & $2.1(2)\times10^{-19}$ & 334.0 & 1 & 5.2&5.2 \\
J1801$-$3210 & 358.92 & $-$4.58 & 7.45 & $-4(4)\times10^{-23}$ & 177.7 & 4, 5 & 6.1&4.0 \\
J1804$-$2858 & 1.99 & $-$3.50 & 1.49 & -- & 232.4 & 6, 7 & 8.2&4.9\\
\hline
\end{tabular}
\begin{tabular}{l}
$^\dagger$PSRs~J1746$-$2758 and J1757$-$2745 were previously published as J1746$-$27 and J1757$-$27, respectively, due to lack of precision in\\ their positions.\\
$^{\dagger\dagger}$Note that YMW16 and NE2001 used the distance of the Sun from the GC as $8.3$~kpc and 8.5~kpc, respectively. However, the\\ systematic offset introduced by this difference in the distance measurement is much smaller than the uncertainty of these electron\\ density models.\\
\\
{\it References}: (1) \citet{ncb+15}; (2) \citet{kek+13}; (3) \citet{mic13}; (4) \citet{bbb+11}; (5) \citet{nbb+14}; \\
(6) \citet{mbc+19}; (7) Barr et al.~(in preparation)
\end{tabular}
\end{center}
\end{table*}

The paper is organized as follows. In \S~\ref{discovery}, we describe the discovery observations of PSR~J1753$-$2819. In \S~\ref{obs}, we present our observations and data processing of all six pulsars. We then discuss the timing analysis of these pulsars in \S~\ref{timing} and report their negative/small observed $\dot P$ measurements. We analyse their dynamics in \S~\ref{dynamics}, including the LOS accelerations combining the Galactic mass distribution and pulsar velocity models (\S~\ref{los_acc} and \ref{galactic_model}), leading to limits on their intrinsic $\dot P$ values (\S~ \ref{pdot}) and their possible distances (\S~\ref{location}). We compare our dynamically-derived distances with DM-derived distances obtained using electron density models in \S~\ref{comparison}. Finally in \S~\ref{concl}, we discuss our results and present conclusions.

\section{Discovery of PSR~J1753$-$2819}
\label{discovery}

The Parkes Multibeam Pulsar Survey \citep[PMPS --][]{mlc+01} has significantly expanded the known pulsar population, resulting in over 800 new pulsar discoveries to-date \citep{mhl+02,kbm+03,hfs+04,fsk+04,lfl+06,ekl09,emk+10,eklk13,kel+09,mlb+12,kek+13}, including 21 millisecond pulsars and 30 rotating radio transients. The PMPS was carried out as a blind pulsar survey along the Galactic plane with $|b|<5^\circ$ using the 13-beam multibeam receiver (with a centre frequency of 1374~MHz and a bandwidth of 288~MHz) on the Parkes radio telescope (PKS) in Australia. Due to advanced and improved pulsar search pipelines being applied to the archival PMPS data, and independent visual inspection of the candidate output plots, five new millisecond pulsars were discovered and reported by \citet{mlb+12}.

These methods also resulted in the discovery of PSR~J1753$-$2819 \citep[originally announced by][]{mic13}. As described by \citet{mlb+12}, the PMPS data were initially dedispersed  using \texttt{dedisperse\_all}\footnote{\url{https://github.com/swinlegion/sigproc}}, a multi-threaded dedisperser compatible with the \textsc{sigproc} software package \citep{lor11}. The resulting time series were then searched using \texttt{seek}\footnote{\url{http://sigproc.sourceforge.net}}. For statistically significant candidates, the raw data were folded at the candidate period using \texttt{prepfold} from the \textsc{presto} software package\footnote{\url{https://www.cv.nrao.edu/~sransom/presto}} \citep{ran11}. The  number of candidates was reduced by selecting those within a parameter space from which pulsars were likely to have been previously missed, i.e.~spin period $P < 50$~ms, DM~$> 10$~pc~cm$^{-3}$. The resulting folded candidates were viewed by eye and  confirmation observations were performed and were  reported by \citet{mlb+12}. This search was undertaken as part of a larger search for fast radio bursts, which classified single pulses from known pulsars detected in the PMPS \citep{mmm+18}. Like the five pulsars reported in \citet{mlb+12}, PSR~J1753$-$2819 was probably missed in earlier searches of the PMPS due to its high DM, short spin period, and binary nature, as well as the large number of candidates present in those searches. Further details of the search are reported in \citet{mlb+12}.

\section{Observations and data processing}
\label{obs}

We observed PSRs~J1748$-$3009, J1746$-$2758, J1757$-$2745, J1801$-$3210, and J1804$-$2858 using the Lovell Telescope (LT) at the Jodrell Bank Observatory in the UK approximately monthly or semi-monthly since their discoveries. After confirming the discovery of PSR~J1753$-$2819, it was observed using the LT approximately every 17 days. All these observations were carried out at `L band' at a centre frequency of 1532~MHz and a bandwidth of 400~MHz. The data were recorded using the `ROACH' pulsar backend \citep{bjk+16}. The observation details are given in Table~\ref{obs_info}. In addition to the ROACH data sets, PSR~J1801$-$3210 was observed using the LT with the `DFB' backend between February 2010 and April 2011. The `DFB' observations were carried out at a slightly different centre frequency of 1520~MHz and a bandwidth of 384 MHz. We used both data sets of this pulsar in the analysis to extend its time baseline. 

In addition to the LT observations, PSRs~J1801$-$3210 and J1804$-$2858 were observed using the Parkes radio telescope since their discoveries. These observations were also taken at `L band' with a centre frequency of 1369~MHz and a bandwidth of 400~MHz and recorded with the CASPSR backend (see Table~\ref{obs_info}). We noticed that some of the calibration observations obtained from PKS as a part of PMPS in the period between December 1998 and August 2001 happened to overlap with the position of PSR~J1748$-$3009, that is, prior to its discovery. Therefore, we used these data and folded them using the pulsar ephemeris and included the resultant detections in our timing analysis. These observations were carried out at 1374~MHz with a bandwidth of 288~MHz and recorded using the `analogue filter bank' (AFB) backend. When combined with the existing data set this extended the observation baseline of PSR~J1748$-$3009 by more than 13 years, we were able to significantly improve the timing measurements while fitting for new timing model parameters (see \S~\ref{timing}).

\begin{table*}
\begin{center}
\caption{
The observation details of pulsars in this analysis. From left to right, we list the pulsar name, telescope used, centre observing frequency and bandwidth, number of bins in the integrated pulse profile and number of observations.
 }
\label{obs_info}
\begin{tabular}{lccccccc}
\hline
\multicolumn{1}{c}{PSR} &
\multicolumn{1}{c}{Telescope} &
\multicolumn{1}{c}{Centre Freq} &
\multicolumn{1}{c}{Bandwidth} &
\multicolumn{1}{c}{Phase bins} &
\multicolumn{1}{c}{Data span} &
\multicolumn{1}{c}{No. of} \\
\multicolumn{1}{c}{ } &
\multicolumn{1}{c}{Backend} &
\multicolumn{1}{c}{(MHz)} &
\multicolumn{1}{c}{(MHz)} &
\multicolumn{1}{c}{ } &
\multicolumn{1}{c}{ } &
\multicolumn{1}{c}{observations} \\
\hline
\hline
J1746$-$2758 & LT/ROACH & 1532 & 400 & 1024 & $12/2013-4/2018$ & 82 \\
J1748$-$3009 & LT/ROACH & 1532 & 400 & 1024 &  $5/2012-11/2018$ & 146 \\
			& PKS/AFB & 1374 & 288 & 32 & $12/1998-8/2001$ & 23 \\
J1753$-$2819 & LT/ROACH & 1532 & 400 & 1024 & $9/2012-10/2018$ & 94 \\ 
J1757$-$27 & LT/ROACH & 1532 & 400 & 1024 & $3/2014-10/2018$ & 80 \\
J1801$-$3210 & LT/ROACH & 1532 & 400 & 1024 & $4/2011-10/2018$ & 74 \\
			& LT/DFB & 1520 & 384 & 1024 & $7/2009-4/2011$ & 60 \\
			& PKS/CASPSR & 1369 & 400 & 256 & $7/2010-4/2018$ &  53 \\
J1804$-$2858 & LT/ROACH & 1532 & 400 & 256 & $12/2013-3/2018$ & 46 \\
			& PKS/CASPSR & 1369 & 400 & 256 & $12/2013-4/2018$ & 55 \\
\hline
\end{tabular}
\end{center}
\end{table*}

We processed these data using the pulsar analysis software package \textsc{psrchive}\footnote{\url{http://psrchive.sourceforge.net}}\citep{hvm04,vdo12,vdk11}. We folded each observation for its entire length and summed all the frequency channels together to form the integrated pulse profile. A time-of-arrival (TOA)  was generated by cross-correlating each of the integrated pulse profiles with a noise-free profile template \citep{tay92} using \texttt{pat}\footnote{we used the FDM (Fourier domain with Markov chain Monte Carlo) method.} in \textsc{psrchive}. This resulted in 83, 169, 98, 75, 187, and 101 TOAs for PSRs~J1746$-$2758, J1748$-$3009, J1753$-$2819, J1757$-$2745, J1801$-$3210, and J1804$-$2858, respectively.

\begin{figure}
\includegraphics[width=8cm]{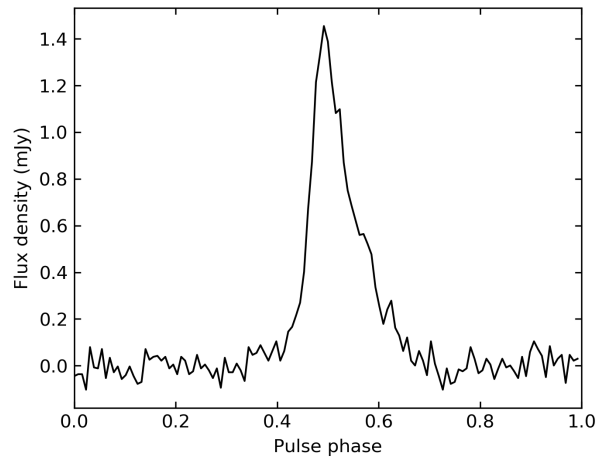}
\centering
\caption{
Integrated pulse profile of recently discovered PSR~J1753$-$2819. Note that the y-axis is scaled in flux density. The mean flux density is estimated to be 0.14~mJy and the pulse widths at 50\% and 10\% of the peak are estimated to be $W_{\rm 50}=1.51$~ms and $W_{\rm 10}=3.98$~ms, respectively.
}
\label{J1753_prof}
\end{figure}

Fig.~\ref{J1753_prof} shows the integrated pulse profile of PSR~J1753$-$2819 as it was not published in the original announcement \citep{mic13}. We estimate the flux density of the pulsar using the radiometer equation \citep[see][]{lk05}, with the system temperature of $T_{\rm sys}=42$~K (i.e.~$T_{\rm sky}=12$~K in the direction of the pulsar according to \citet{hssw82} with a spectral index of $-2.6$ and $T_{\rm rec} + T_{\rm spill} = 30$~K) and the telescope gain of 0.9~K/Jy, and scale the profile accordingly. These calculations estimate the mean flux density of the pulsar to be 0.14~mJy. We further measure the pulse width, at 50\% and 10\% of the peak to be $W_{\rm 50}=1.51$~ms and $W_{\rm 10}=3.98$~ms, respectively.

\section{Timing pulsars}
\label{timing}

We fit timing models including astrometric and rotational frequency parameters for PSRs~J1746$-$2758, J1748$-$3009, J1753$-$2819, J1757$-$2745, J1801$-$3210, and J1804$-$2858 to the TOAs. The timing solutions are improved by minimising the $\chi^2$ value of the timing residuals between the observed and model-predicted TOAs using the pulsar timing software \textsc{tempo2} \citep{ehm06,hem06,he12}. Since we do not have multi-frequency observations, we keep the DM of the pulsar fixed at its previously published value  (see Table~\ref{psr_info}). Note that PSRs~J1748$-$3009, J1753$-$2819, and J1801$-$3210 are in binary orbits and thus, we include Keplerian parameters in their timing models \citep[see][]{lk05} using the binary model `BT' \citep{bt76}. When TOAs from different telescopes/backends are available (see Table~\ref{obs_info}), we combined them by fitting for a time offset (`JUMP') in the timing model of the pulsar to take into account any systematic delays between the data sets. During the fitting process, the topocentric TOAs are converted to the Solar-system barycentric coordinate time (TCB) using the DE436\footnote{This new solar-system ephemeris DE436 is based on \citet{fwb+14}.} ephemeris \citep{fwb09} and placed on the Terrestrial Time standard BIPM2015\footnote{This time standard has been obtained according to principles given in \citet{gui88} and \citet{pet03b}.}. The resultant timing parameters of our pulsars are given in Table~\ref{sol_all_1} and \ref{sol_all_2}. 

\begin{table*}
\begin{center}
\caption{
The timing model parameters of the newly discovered PSR~J1753$-$2819 and the updated timing model parameters of  PSRs~J1746$-$2758, and J1748$-$3009. Due to the lack of multi-frequency observations, we present the DM values (and their uncertainties, if available) reported in their discovery analyses and kept them fixed. The 1$\sigma$ uncertainties of measurements are given in parentheses. The characteristic age ($\tau=P/2\dot{P}$) and the surface magnetic field ($B_S=3.2\times10^{19}\sqrt{P\dot{P}}$~G) are computed based on the intrinsic $\dot P$ limits estimated using the dynamics of each
pulsar as described in \S~\ref{pdot}.
}
\label{sol_all_1}
\begin{tabular}{lccc}
\hline
\multicolumn{1}{l}{Timing parameter} &
\multicolumn{1}{c}{J1746$-$2758} &
\multicolumn{1}{c}{J1748$-$3009} &
\multicolumn{1}{c}{J1753$-$2819} \\
\hline
Data span (MJD)  & 56639 -- 58423 & 51148 -- 58428 & 56174 -- 58423  \\
Number of TOAs & 82 & 169 &  94 \\
Weighted rms timing residual ($\upmu$s) & 545 & 86 & 82 \\
Reduced $\chi^2$ value & 1.00 & 0.99 & 0.99  \\
Units & TCB & TCB & TCB \smallskip \smallskip \\ 
Measured parameters: & \smallskip \\
Right ascension RA (J2000) & 17:45:52.248(3) & 17:48:23.7582(3) & 17:53:56.1837(4)   \\
Declination Dec. (J2000) & $-$27:58:38.0(5) & $-$30:09:11.28(4) & $-$28:19:30.03(7)  \\
Spin frequency $f$ ($\rm s^{-1}$) & 2.051164249099(9) & 103.26352668931(4) & 53.69521954620(1) \\
Spin frequency $\rm 1^{st}$ derivative $\dot f$ ($\rm s^{-2}$) & $-1.86(2)\times10^{-17}$ & $3.817(2)\times10^{-16}$ & $1.423(4)\times10^{-16}$ \\
Spin frequency $\rm 2^{nd}$ derivative $\ddot f$ ($\rm s^{-3}$) & $-4.9(19)\times10^{-26}$ & $-2.2(3)\times10^{-26}$ & --  \\
Reference epoch (MJD) & 57531 & 54788 & 57299  \\
Dispersion measure DM (cm$^{-3}$~pc) & 422(9) & 420.2 & 298 \\
Orbital period $P_{\rm b}$ (d) & -- & 2.93382034(2) & 0.387677373(3) \\
Epoch of periastron $T_o$ (MJD) & -- & 56999.183308(4) & 56999.821829(4)  \\
Projected semi-major axis $x$ (lt-s) & -- & 1.320085(9) & 0.20967(1)  \\
Orbital period derivative $\dot{P_{\rm b}}$ (s/s) & -- & $-2.6(10)\times10^{-11}$ & --  \smallskip \smallskip \\
Derived parameters: \smallskip \\
Gal. longitude $l$ (deg) & 0.8473(1) & 359.272301(8) & 1.46179(2)  \\
Gal. latitude $b$ (deg) & 0.45243(8) & $-1.147126(6)$ & $-1.25174(1)$ \\
Spin period $P$ (s) & 0.487527998033(2) & 0.009683961337179(4) & 0.018623631832617(5) \\
Spin period derivative $\dot {P}$ (s/s) & $4.42(5)\times10^{-18}$ & $-3.580(2)\times10^{-20}$ & $-4.94(2)\times10^{-20}$  \smallskip\\
Intrinsic $\dot {P}$ (s/s) & $<$$1.48\times10^{-17}$ & $<$$1.38\times10^{-19}$ & $<$$2.08\times10^{-19}$  \\
Characteristic age $\tau$ (Gyr) & $>$0.5 & $>$1.1 & $>$1.4  \\
Surface magnetic field $B_s$ ($\times 10^{9}$~G) & $<$85 & $<$1.1 & $<$2.0 \\
Companion mass (M$_\odot$) & -- & $0.08-0.20$ & $0.05-0.12$ \\
\hline
\end{tabular}
\end{center}
\end{table*}

\begin{table*}
\begin{center}
\caption{
Same as Table~\ref{sol_all_1}, but for PSRs~J1757$-$2745, J1801$-$3210, and J1804$-$2858. 
}
\label{sol_all_2}
\begin{tabular}{lccc}
\hline
\multicolumn{1}{l}{Timing parameter} &
\multicolumn{1}{c}{J1757$-$2745} &
\multicolumn{1}{c}{J1801$-$3210} &
\multicolumn{1}{c}{J1804$-$2858} \\
\hline
Data span (MJD)  &  56728 -- 58422 & 55032 -- 58423 & 56638 -- 58213 \\
Number of TOAs &  80 & 187 & 101 \\
Weighted rms timing residual ($\upmu$s) & 12 & 28 & 8 \\
Reduced $\chi^2$ value & 0.98 & 0.99 & 0.97 \\
Units & TCB & TCB & TCB \smallskip \smallskip \\ 
Measured parameters: & \smallskip \\
Right ascension RA (J2000) &  17:57:54.7826(1) & 18:01:25.8872(1) & 18:04:01.52323(4) \\
Declination Dec. (J2000) &  $-$27:45:40.16(2) & $-$32:10:53.733(9) & $-$28:58:46.608(7)\\
Proper motion in RA, PMRA (mas/yr) & $-11.4(11)$ & $-4.0(6)$ & $-6.4(5)$ \\
Proper motion in Dec., PMDEC (mas/yr) & 54(17) & $-4(3)$ & $-11(5)$ \\
Spin frequency $f$ ($\rm s^{-1}$) & 56.538013330516(8) & 134.16363857905(1) & 669.93358254482(5) \\
Spin frequency $\rm 1^{st}$ derivative $\dot f$ ($\rm s^{-2}$) & $4.17(5)\times10^{-17}$  & $-1.3(2)\times10^{-18}$ & $3.399(9)\times10^{-16}$ \\
Spin frequency $\rm 2^{nd}$ derivative $\ddot f$ ($\rm s^{-3}$) & $-4.7(18)\times10^{-26}$ & $-2.5(5)\times10^{-26}$ & $-2.6(9)\times10^{-25}$ \\
Spin frequency $\rm 3^{rd}$ derivative $\dddot f$ ($\rm s^{-4}$) & $-5.1(14)\times10^{-33}$ & -- & -- \\
Reference epoch (MJD) & 57575 & 56635 & 57425 \\
Dispersion measure DM (cm$^{-3}$~pc) & 334 & 177.713(4) & 232.4 \\
Orbital period $P_{\rm b}$ (d) & -- & 20.77169953(3) & --\\
Epoch of periastron $T_o$ (MJD) & -- & 56518.268546(1) & --\\
Projected semi-major axis $x$ (lt-s) & -- & 7.809320(3) & --\\
Derived parameters: \smallskip \\
Gal. longitude $l$ (deg) & 2.390743(5) & 358.921963(2) & 1.995292(2) \\
Gal. latitude $b$ (deg) & $-1.726491(3)$ & $-4.577230(2)$ & $-3.497440(1)$\\
Spin period $P$ (s) & 0.017687215045108(3) & 0.0074535843734652(5) & 0.0014926852841164(1) \\
Spin period derivative $\dot {P}$ (s/s) & $-1.30(2)\times10^{-20}$  & $7.4(8)\times10^{-23}$ & $-7.57(2)\times10^{-22}$ \smallskip\\
Intrinsic $\dot {P}$ (s/s) & $<$$1.42\times10^{-19}$ & $<$$3.50\times10^{-20}$ & $<$$7.92\times10^{-21}$  \\
Characteristic age $\tau$ (Gyr) & $>$2.0 & $>$3.4 & $>$3.0 \\
Surface magnetic field $B_s$ ($\times 10^{9}$~G) & $<$1.6 & $<$0.5 & $<$0.1 \\
Companion mass (M$_\odot$) & -- & $0.14-0.35$ &  \\
\hline
\end{tabular}
\end{center}
\end{table*}

\begin{table*}
\begin{center}
\caption{
Summary of the parameters of pulsars used in our dynamic analysis. The angular separation $\theta$ of the pulsar from the GC is calculated based on the Galactocentric coordinates $(l, b)$. The uncertainties of $l$, $b$, and $\theta$ are estimated based on the uncertainties of pulsar positions measured from timing (Table~\ref{sol_all_1} and \ref{sol_all_2}). The projected separation $r_\perp$ of the pulsar from the GC in the sky plane is calculated based on a GC distance of $8.2\pm0.1$~kpc from the Sun \citep{bst+19}. Note that the uncertainty in $r_\perp$ is dominated by the uncertainty in the GC distance.
 }
\label{summary}
\begin{tabular}{lccccc}
\hline
\multicolumn{1}{c}{PSR} &
\multicolumn{1}{c}{$l$} &
\multicolumn{1}{c}{$b$} &
\multicolumn{1}{c}{$\theta$} &
\multicolumn{1}{c}{$r_\perp$} &
\multicolumn{1}{c}{$\dot{P}_{\rm obs}/P$} \\
\multicolumn{1}{c}{ } &
\multicolumn{1}{c}{($^\circ$)} &
\multicolumn{1}{c}{($^\circ$)} &
\multicolumn{1}{c}{($^\circ$)} &
\multicolumn{1}{c}{(pc)} &
\multicolumn{1}{c}{($\times10^{-10}$~yr$^{-1}$)} \\
\hline
\hline
J1746$-$2758 & 0.8473(1) & 0.45243(8) & 0.9605(1) & 137(2) & $2.86(3)$ \\
J1748$-$3009 & 359.272301(8) & $-$1.147126(6) & 1.358445(7) & 194(2) & $-1.1657(6)$ \\
J1753$-$2819 & 1.46176(2) & $-$1.25174(1) & 1.92438(1) & 275(3) & $-0.836(3)$ \\
J1757$-$2745 & 2.390743(5) & $-$1.726491(3) & 2.94868(5) & 422(5) & $-0.233(3)$ \\
J1801$-$3210 & 358.921963(2) & $-$4.577230(2) & 4.702204(2) & 672(7) & $0.0031(4)$ \\
J1804$-$2858 & 1.995292(2) & $-$3.497440(1) & 4.025958(1) & 576(6) & $-0.1600(4)$\\
\hline
\end{tabular}
\end{center}
\end{table*}

\subsection{PSR~J1748$-$3009}
\label{J1748}

This binary millisecond pulsar has a spin period of $9.7$~ms and is located with an angular separation of $1\fdg36$ from the GC (see Table~\ref{summary}). Based on about 20 years of our timing data, by combining the early PKS observations with LT observations, the timing parameters of the pulsar improved significantly. The timing model constrains the $\dot P$ measurement of the pulsar for the first time to be $-3.580(2)\times10^{-20}$~s/s. This negative value indicates that the measured $\dot P$ is not entirely intrinsic, rather induced by the acceleration of the pulsar due to the Galactic potential, and we constrain the limit on the intrinsic value based on dynamics in \S~\ref{pdot}. We further find that the timing solution of the pulsar requires a second time derivative of the rotational frequency ($\ddot{f}$), with an approximately 7$\sigma$ significance (see Table~\ref{sol_all_1}), to achieve white residuals. Our timing analysis also places a limit on the orbital period time derivative $\dot{P}_b$ of the pulsar with a significance of 2.6$\sigma$. We use this limit independently to further constrain the intrinsic period derivative of the pulsar in \S~\ref{pdot}.

\subsection{PSR~J1753$-$2819}

The newly discovered PSR~J1753$-$2819 \citep{mic13} is a 18.6~ms mildly-recycled pulsar in a 9.3-hr binary orbit (see Table~\ref{sol_all_1}) with a DM of 298~cm$^{-3}$~pc. It is located at an angular separation of $1\fdg92$ from the GC, resulting in a projected separation of 275~pc in the sky-plane at the distance of the GC (see Table~\ref{summary}). Initial timing observations during the discovery showed a mild evidence for a negative $\dot P$, and the current value is measured to be $-4.94(2)\times10^{-20}$~s/s based on our 6 years of LT observations. This negative $\dot P$ value indicates that this measurement is also affected by the pulsar dynamics in the Galactic potential (see \S~\ref{pdot}). 

\subsection{PSR~J1757$-$2745}

PSR~J1757$-$2745 is an isolated mildly-recycled pulsar with a spin period of 17.7~ms and located at an angular separation of $2\fdg95$ from the GC (see Table~\ref{summary}). Our 4.5 years of observations improved its timing solution, updating the $\dot P$ measurement to be $-1.30(2)\times10^{-20}$~s/s (see Table~\ref{sol_all_2}). This negative value indicates that this $\dot P$ measurement is also affected by the pulsar dynamics in the Galactic potential, and we study this in detail (see \S~\ref{pdot}). The timing analysis also places a limit on the proper motion, particularly in Right Ascension (see Table~\ref{sol_all_2}).

The previous timing analysis of this pulsar reported a $\dot P$ measurement of $2.1(2)\times10^{-19}$~s/s \citep{ncb+15} based on a short time span of 206 days. However, we now confirm that this value is not correct due to the short data span and resultant large covariance between the rotational and astrometric parameters. 

\subsection{PSR~J1801$-$3210}
\label{J1801}

PSR~J1801$-$3210 has a spin period of 7.45~ms and is located at an angular separation of $4\fdg7$ from the GC (see Table~\ref{summary}). 
As shown in Table~\ref{sol_all_2} based on our 9 yr data set, we find improved timing parameters of the pulsar, indicating that $\dot P$ measurement is constrained to be a positive value of $7.4(8)\times10^{-23}$~s/s. This measurement is still smaller than the lowest positive $\dot P$ measured to-date for any known pulsar\footnote{\url{http://www.atnf.csiro.au/people/pulsar/psrcat}}.

The previous timing analysis of this pulsar showed a low-significance $\dot P$ measurement of $-4(4)\times10^{-23}$~s/s based on a 4~yr long data set \citep{nbb+14}. They argued that the pulsar is possibly located behind the sky-plane which passes through the GC along the LOS. In that case the negative Galactic acceleration affects the positive intrinsic spin-down of the pulsar and results in an extremely low measured $\dot P$ value (see \S~\ref{los_acc} for details). Although the DM-derived distance from the NE2001 electron density model places the pulsar in front of the sky-plane (see Table~\ref{psr_info}), they argued that the electron density models can contain large uncertainties. Therefore, our exceptionally small $\dot P$ measurement indicates the scenario of the pulsar being located behind the sky-plane through the GC proposed by \citet{nbb+14} is still valid (which we will show independently in \S~\ref{location}).

\subsection{PSR~J1804$-$2858}
\label{J1804}

PSR~J1804$-$2858 is a solitary millisecond pulsar and has the third lowest spin period known to-date, a spin period of $1.49$~ms \citep[][Barr et al., {\it in prep}]{mbc+19}. It is located at an angular radius of 4$^\circ$ from the GC (Table~\ref{summary}) and the DM is 232~cm$^{-3}$~pc, which places the pulsar at a distance of 8.2~kpc from the Earth based on the YMW16 electron density model (see Table~\ref{psr_info}). The discovery analysis of the pulsar did not have a $\dot{P}$ measurement, however our 4.3 years of timing data measures a significant $\dot{P}$ value of $-7.57(2)\times10^{-22}$~s/s (see Table~\ref{sol_all_2}). As mentioned above for other pulsars, this measured $\dot{P}$ value is not entirely intrinsic, rather induced by the acceleration of the pulsar along its LOS due to the Galactic potential. We study its dynamics in detail combining with the Galactic mass distribution and constrain the intrinsic $\dot{P}$ value in \S~\ref{pdot}.     

\subsection{PSR~J1746$-$2758}
\label{J1746}

This isolated $0.49$-s spin period pulsar has an angular separation of just $0\fdg96$ from the GC (see Table~\ref{summary}). Our updated timing solution based on more than four years of observations indicates a $\dot P$ measurement for the first time to be $4.42(5)\times10^{-18}$~s/s (see Table~\ref{sol_all_1}). Comparing this measurement with $\dot P$ values of known pulsars that have a similar pulse period, it is evident that this value is slightly smaller than the majority, but within the distribution. On the other hand, one can imagine, given the pulsar's location, that it is closer to the GC than the DM-derived distance (see Table~\ref{psr_info}) and thus, the measured $\dot P$ is not entirely intrinsic, but rather induced by the Galactic potential. However, such a claim is not likely to be straightforward. Further note that the maximum Galactic acceleration in the direction of the pulsar is greater than the measured $\dot{P}/P$ and thus, we study its dynamics due to the Galactic potential in \S~\ref{pdot} in detail.

\section{Dynamics of pulsars}
\label{dynamics}

The timing models for PSRs~J1748$-$3009, J1753$-$2819, J1757$-$2745, and J1804$-$2858 show negative observed $\dot{P}$ measurements (see \S~\ref{timing}), indicating that these values are not entirely intrinsic and have a dynamic contribution. We also note that the observed $\dot P$ of PSR~J1801$-$3210 is extremely small, so that it is also likely to be affected by dynamics (see \S~\ref{J1801}). In this section, we model the observed unusual $\dot P$ measurements as being due to acceleration of these pulsars along their LOSs, and then constrain their intrinsic $\dot P$ values and possible distances.

\subsection{Pulsar line-of-sight acceleration}
\label{los_acc}

As noted in previous studies \citep[e.g.][]{dt91,nt95,mnf+16,dcl+16, psl+17a}, the observed $\dot{P}_{\rm obs}/P$ of the pulsar is a combination of individual terms and can be expressed as
\begin{equation}
\label{acc_eq}
\frac{\dot{P}_{\rm obs}}{P} = \frac{\dot{P}_{\rm int}}{P} + \frac{a}{c} + \frac{a_{\rm \mu}}{c},
\end{equation}
where, $\dot{P}_{\rm obs}$ and $\dot{P}_{\rm int}$ are the observed and intrinsic spin period derivatives, respectively, $a$ is the LOS acceleration of the pulsar with respect to the Sun due to the 3-dimensional Galactic potential, $a_{\rm \mu}$ is the acceleration due to transverse motion of the pulsar \citep[i.e. the Shklovskii effect --][]{shk70}, and $c$ is the speed of light. In previous studies, $a/c$ is computed as two separate components, namely the parallel component in the Galactic plane and the vertical component towards the Galactic plane \citep[see][]{dt91,nt95,bjs+16,dcl+16,lwj+09}. These components were estimated separately based on Galactic rotation curve models and Galactic vertical acceleration models for a given $(l,b)$ of a pulsar using the Galactocentric distance of the solar system ${\rm R}_\odot$, the circular rotational speed of the Galaxy at ${\rm R}_\odot$, and the pulsar distance. Since the location of the Sun is only $25\pm5$~pc above the Galactic plane \citep{bg16,bst+19}, its vertical acceleration towards the plane is very small and neglected in general \citep[e.g.][]{dt91}. In contrast, we directly use 3-dimensional Galactic mass distribution models (see \S~\ref{galactic_model}) and thus, estimate the 3-dimensional acceleration of the pulsar and the Sun separately at their locations in the Galaxy. We then use these acceleration terms to compute the relative acceleration $a/c$ of the pulsar along its LOS with respect to the Sun. We note that the acceleration of the Sun is insignificant compared to that of the pulsar when the pulsar is located near the GC. We describe the Galactic mass distribution models used in this study and the estimation of $a/c$ in \S~\ref{galactic_model}.

The Shklovskii effect can be written as $a_{\rm \mu}/c = v_T^2/cd$ $(\equiv \mu^2 d/c)$, where $v_T$ and $\mu$ are the transverse velocity and the proper motions of the pulsar with respect to the Sun, respectively, and $d$ is the pulsar distance \citep{shk70}. Our timing solutions do not measure significant proper motions of the pulsars considered here, rather place limits (see Table~\ref{sol_all_1} and \ref{sol_all_2}). Thus, we assume typical pulsar properties to estimate this term. We note that the velocity of a pulsar is an intrinsic property of the source and the proper motion is an observed measurement that depends on the pulsar distance and the relative motion of the pulsar and the Sun. Therefore, we use typical pulsar 3-dimensional peculiar velocities and then estimate the expected $a_{\rm \mu}/c$. In general, young pulsars have higher 3-dimensional velocities compared to millisecond pulsars \citep{lml+98}. Therefore, we assume a typical value of 100~km/s, which is consistent with the millisecond pulsar population \citep[e.g.][]{gsf+11,mnf+16}, and 320~km/s, which is consistent with the ordinary pulsar population \citep[e.g.][]{hllk05,vic17}, in this study. We describe our calculation of $a_{\rm \mu}/c$ in Appendix~\ref{velocity} in detail. We note that if the actual velocity of the pulsar is small, then the $a_{\rm \mu}/c$ contribution to the observed $\dot{P}_{\rm obs}/P$ becomes insignificant, and vice-versa (see Eq.~\ref{acc_eq}).

\begin{figure}
\includegraphics[width=8cm]{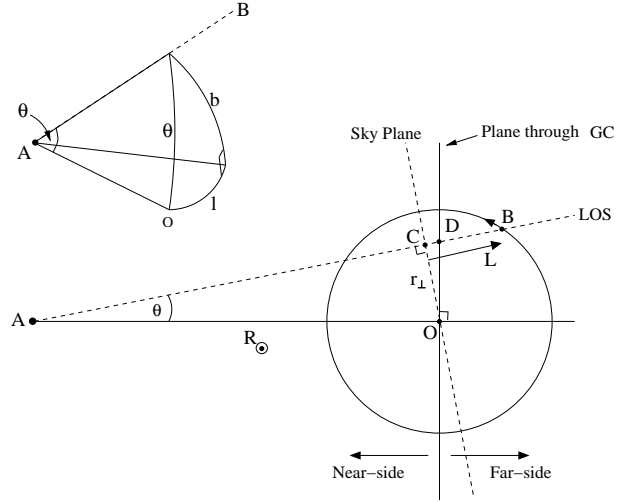}
\centering
\caption{
Schematic diagram of the geometry used in this analysis. The GC and the observer are denoted as O and A, respectively. The pulsar is denoted by point B and the sky-plane to the pulsar is defined as the plane which passes through point C and the GC normal to the LOS. The projected distance $r_\perp$ ($=R_{\odot}\sin(\theta)$) of the pulsar is denoted by OC. The distance $L$ is measured from the sky-plane along the LOS. The angular separation of the pulsar $\theta$ from the GC is measured using the Galactic longitude $l$ and latitude $b$ through spherical trigonometry. Note that the Galactic acceleration becomes negative (or positive) along the LOS when the pulsar is located behind (or in front of the) the sky-plane. We define the `far-side' and the `near-side' of the Galaxy based on the plane that passes through the GC normal to the line joining A and O. The distance to the pulsar from the Sun is $R_{\odot}\cos(\theta)+L$. The distance CD is estimated as $r_\perp \tan(\theta)$.
}
\label{diagram}
\end{figure}

The Galactic term $a/c$ has both positive and negative signs depending on the source location along the LOS in the Galaxy. When the source is located behind (or in front of) the sky-plane which passes through the GC, then $a/c$ becomes negative (or positive). We note that the Shklovskii term $a_{\rm \mu}/c$ is always positive. Given that the intrinsic $\dot{P}$ is positive, the pulsar must be located behind the sky-plane in order to explain any observed negative $\dot{P}$, and the observed $\dot{P}/P$ should be dominated by the Galactic contribution (see Eq.~\ref{acc_eq}). A schematic diagram of the geometry used in this study and the definitions are given in Fig.~\ref{diagram}.

\subsection{Galactic mass distribution models and pulsar line-of-sight acceleration}
\label{galactic_model}

We use the Galactic mass distribution model proposed in \citet{sof13,sof17} in our analysis to estimate $a/c$. They analysed high-resolution longitude-velocity diagrams of the Galaxy and constructed the rotation curve including the central black hole. Then they deconvolved the rotation curve and fit several components to the observed velocities across the Galaxy, resulting in two exponential-spherical bulges (i.e.~inner and main bulges), an exponential flat disk, and a Navarro-Frenk-White \citep{nfw96} dark halo. Using this information, \citet{sof13} modelled the volume mass density of each component and then the mass distribution of the Galaxy  \citep[see Table~6 in ][for updated values]{sof17}. This study used observations covering the inner region (from $\sim$1~pc) of the Galaxy and thus, required several bulge components to appropriately model the data.

For comparison, we also use the Galactic mass distribution model proposed by \cite{mcm17}. This model uses a main bulge, two stellar discs, two gas discs, and a dark halo. The \citet{mcm17} model is an improved version of their previous model \citep{mcm11} which now includes gas discs and uses new observations of maser sources in the Galaxy. We note that these models mainly focus on the region beyond the bulge (where $r_\perp > 1$~kpc) in the Galaxy. Since our pulsars are likely located around the central region of the Galaxy (within a $r_\perp < 1$~kpc, see Table~\ref{summary}), we use the \citet{sof13} model with updated parameters given in \citet{sof17} as our main Galactic mass distribution model in the analysis.

Throughout the analysis, we assume the Galactocentric distance of the solar system to be ${\rm R}_\odot = 8.2\pm0.1$~kpc and the central black hole mass to be $(4.2\pm0.2)\times10^6$~M$_\odot$ \citep{bst+19,vas19,bg16,sof13,sof17,mcm17}. These values are consistent with those derived in previous studies within their uncertainties \citep[e.g.][]{gpe+00,gsh+05,gsw+08,get+09,rmz+09,hna+12,sch12,brm+11,rmb+14}. We further note that this ${\rm R}_\odot$ is broadly consistent with the recently measured GC distance, which has an uncertainty of only $0.3$\% \citep{aab+19}.

To be consistent with the above Galactic models \citep{sof13,mcm17}, we define the volume mass density of the bulge component in cylindrical Galactocentric coordinates 
\begin{equation}
\label{bulge}
\rho_{\rm b}(R,z) = \frac{\rho_{\rm 0,b}}{(1+r/r_0)^\alpha}\exp[-(r/r_{\rm cut})^\zeta],
\end{equation}
where, $\rho_{\rm 0,b}$ is the central mass density and $r=\sqrt{R^2 + (z/q)^2}$ with the axis ratio $q$. 
\citet{sof13} includes two bulges with $\zeta=1$, $\alpha=0$, $r_0 = 1$~kpc, and $q=1$ (i.e.~spherically symmetric assumption), while \citet{mcm17} includes a single bulge with $\zeta=2$, $\alpha=1.8$, $r_0= 0.075$~kpc, and $q=0.5$ (i.e.~an axially symmetric assumption). All best-fit parameter values in their models are given in Table~\ref{models}. The volume mass density of the stellar disc component,
\begin{equation}
\label{disc}
\rho_{\rm d}(R,z) = \frac{\Sigma_{\rm 0,d}}{(2z_{\rm d})^\eta}  \exp \left( -\frac{|z|}{z_{\rm d}} - \frac{R}{R_{\rm d}}  \right).
\end{equation}
Here, $\Sigma_{\rm 0,d}$ is the central surface density, $z_d$ is the scale height, and $R_d$ is the scale length. \citet{sof13} uses a flat stellar disc approximation (i.e.~$z_d=0$ an $\eta=0$), while \citet{mcm17} uses thin and thick stellar discs with $z_d = 0.3$ and $0.9$~kpc, respectively, and $\eta=1$. The surface density and the scale radius of the two models are given in Table~\ref{models}. In addition to stellar discs, \citet{mcm17} uses two gas discs representing $H_1$ and $H_2$ gas in the Galaxy. These gas discs are defined with mass densities
\begin{equation}
\label{gas}
\rho_{\rm g}(R,z) = \frac{\Sigma_{\rm 0,d}}{4z_{\rm d}}  \exp \left( -\frac{R_{\rm m}}{R} - \frac{R}{R_{\rm d}}  \right) \sech^2 (z/2z_{\rm d}),
\end{equation}
with holes in their centres. The $H_1$ and $H_2$ discs have scale height $z_{\rm d}$ of $0.085$~kpc and $0.045$~kpc, and central hole scale radius $R_{\rm m}$ of 4~kpc and 12~kpc, respectively. In both \cite{sof13} and \citet{mcm17}, the dark matter halo is defined as having mass density
\begin{equation}
\label{halo}
\rho_{\rm h}(R,z) = \frac{\rho_{\rm 0,h}}{X(1+X)^2},
\end{equation}
where $X=\sqrt{R^2 + z^2}/h$ and $\rho_{\rm 0,h}$ and $h$ are the scale density and core radius, respectively. 
The fitting procedure and the best-fit values for the above parameters are given in \citet{sof17} and \cite{mcm17}, and also in Table~\ref{models} of this study.

\begin{table}
\begin{center}
\caption{
The parameters of the main Galactic mass distribution model SOF13 used in this analysis, i.e.~\citet{sof13} with updated values given in \citet{sof17}. This model uses two bulges (i.e.~inner and main), a stellar disc, and a dark halo. For comparison, we use MCM17 in the analysis and its relevant parameters are also given \citep{mcm17}. This model uses only a single bulge with two stellar (i.e. thin and thick) and two gas discs (i.e. $H_1$ and $H_2$), and a dark halo. The central volume ($\rho_{\rm 0,b}$) and surface mass density ($\Sigma_{\rm 0,d}$) parameters are given in units of M$_\odot$/pc$^3$ and M$_\odot$/pc$^2$, respectively, and the radii are given in kpc.     
 }
\label{models}
\begin{tabular}{lccc}
\hline
\multicolumn{1}{c}{Model} &
\multicolumn{1}{c}{Bulge} &
\multicolumn{1}{c}{Disc} &
\multicolumn{1}{c}{Dark halo} \\
\multicolumn{1}{c}{} &
\multicolumn{1}{c}{$\rho_{\rm 0,b}$ , $r_{\rm cut}$} &
\multicolumn{1}{c}{$\Sigma_{\rm 0,d}$ , $R_{\rm d}$} &
\multicolumn{1}{c}{$\rho_{\rm 0,h}$ , h}  \\
\hline
\hline
SOF13 & Inner: $3.7\times10^4$, &  $292$ , 4.9 & $0.029$ , 10 \\
 & $0.0035$ & & \\
 & Main: $2.1\times10^2$, &  &  \\
 & 0.12 & & \\
 \\
MCM17 & $98.4$ , 2.1 & Thin: 896 , 2.5 & 0.00854 , 19.6 \\
 & & Thick: 183 , 3.02 & \\
 & & H$_1$: 53.1 , 7 & \\
 & & H$_2$: 2180 , 1.5 & \\
\hline
\end{tabular}
\end{center}
\end{table}

\begin{figure}
\includegraphics[width=8.3cm]{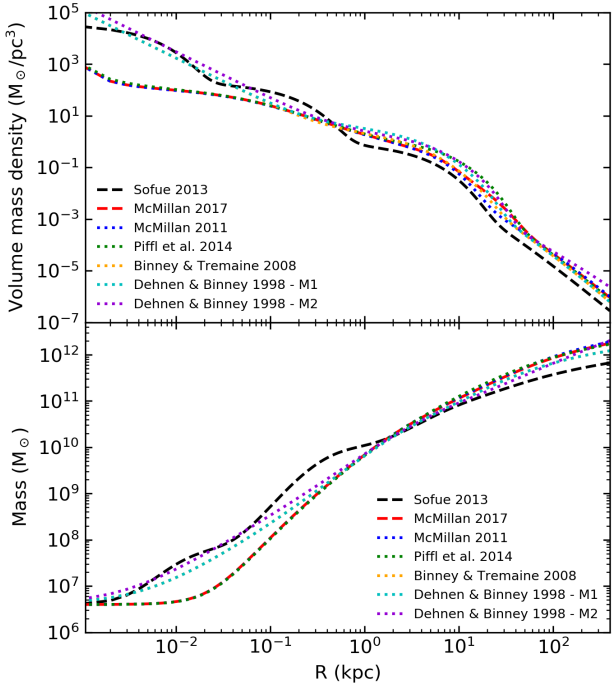}
\centering
\caption{
The volume mass density ({\it top}) and the mass distribution ({\it bottom} -- i.e. the mass interior to radius $R$) of the Galaxy as a function of radius $R$ measured from the GC. The different colours show the different Galactic models used. Note that the \citet{sof13} model with updated values given in \citet{sof17} provides better constraints within the inner Galaxy, in particular in the bulge regions. Other models mainly focus on the regions beyond the bulge and follow similar constraints. The \citet{db98} models were derived for the outer Galaxy and thus, they do not provide meaningful results when $R\lesssim$8~kpc. Since the Galactic central BH is included in all models, their mass distributions converge to the BH mass of $4.2\times 10^6$~M$_\odot$ when $R\rightarrow0$. We use \citet{sof13} as our main model to constrain the properties of our pulsars.  
}
\label{density_mass}
\end{figure}

Using the above components and the central BH, we estimate the total volume mass density of the Galaxy as a function of radius $R$ (see the {\it top} panel in Fig.~\ref{density_mass}). Based on mass density, we then estimate the  mass distribution (i.e. the total mass interior to a given radius $R$) of the Galaxy (see the {\it bottom} panel in Fig.~\ref{density_mass}). For further comparison, we also use previously published Galactic mass distribution models \citep{pbm+14,bt08,db98} with their best fit parameters and calculate the volume mass density and the mass distribution as a function of radius (see Fig.~\ref{density_mass}). It can be seen that all these additional models were focused on the regions beyond the bulge and thus, derive similar variations in their density and mass curves to those derived from \citet{mcm17} model. We note that the Galactic models in \citet{db98} are derived for the outer Galaxy and thus, its density and mass estimates are not meaningful at radii $R\lesssim8$~kpc. As shown in Fig.~\ref{density_mass}, the model derived in \citet{sof13} provides more features and better constraints within the inner region of the Galaxy, particularly in the bulges where our pulsars are likely to be located.

We further note that some Galactic mass models suggest that there is a need for a ``barred bulge'' component in the central region of the Galaxy \citep[e.g.,][]{mc10,nug+10,gf10,nfa+12}. These models are non-axisymmetric and are typically computationally expensive to implement. By comparing the \citet{sof13} model and the bar/bulge model given in \citet{pgwn17}, we noticed that the mass distribution at a given radius of the two models are consistent with each other. Compared to an axially symmetric model, a bar/bulge model can cause a slight difference in the Galactic gravitational potential along the LOS of the pulsar depending on the orientation of the bar component, leading to a slight change in our results. However, this change is within the uncertainties of our estimated $\dot{P}_{\rm int}$ values and pulsar distances. Therefore, we simply use the above mentioned axially symmetric Galactic mass models in this study.

We first calculate the gravitational potential $\Phi$ at the pulsar location $(R, z)$ due to the Galactic mass distribution, and then estimate the Galactic acceleration in $R$ and $z$ directions, namely $-d\Phi/dR$ and $-d\Phi/dz$. To do this, we use the computer software \texttt{GalPot}\footnote{\url{https://github.com/PaulMcMillan-Astro/GalPot}} \citep{mcm17,mcm16}, which computes the Galactic potential and the acceleration for any given Galactic mass distribution model. This software was based on \citet{db98} and the details of the computation are given in \citet{mcm17}. We note that since all these Galactic mass models are axially symmetric (see Eq.~\ref{bulge} -- \ref{halo}), the acceleration does not depend on the azimuthal angle. Once the acceleration components of the pulsar and the Sun at their given locations in the Galaxy in $R$ and $z$ directions are computed, we can estimate the relative LOS acceleration $a/c$ of the pulsar with respect to the Sun (see \S~\ref{los_acc}). The acceleration $a/c$ varies with the distance to the pulsar, providing a negative (or positive) value when the pulsar is located behind (or in front of) the sky-plane.

\begin{figure}
\includegraphics[width=8.4cm]{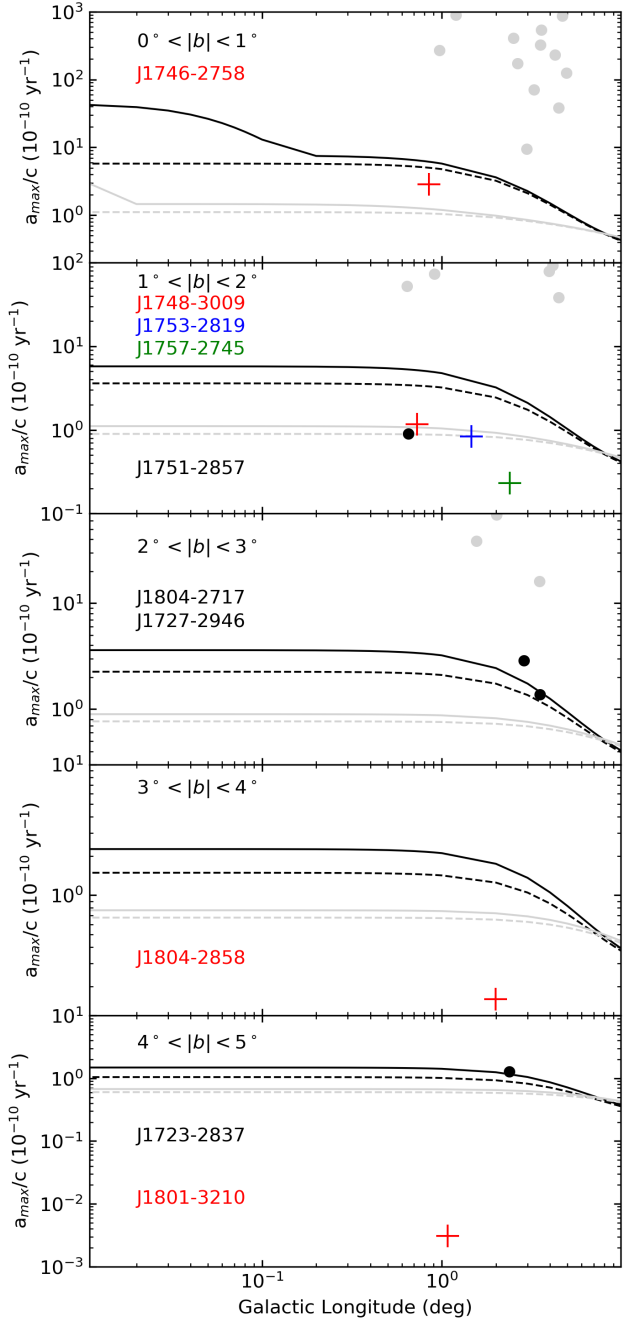}
\centering
\caption{
The estimated maximum LOS acceleration values ({\it black} curves) based on the main Galactic model \citep{sof13} used in our analysis, as a function of Galactic longitude for various Galactic latitudes (see each panel). There are two separate acceleration curves shown in each panel corresponding to the lower ({\it solid}) and the upper ({\it dashed}) bound of the Galactic latitude used. The values of $\dot{P}_{\rm obs}/P$ of pulsars in our study are marked with crosses and the colour-code represents different sources as named in each panel. For clarity, we plot the absolute value of $\dot{P}_{\rm obs}/P$ of PSRs~J1748$-$3009, J1753$-$2819, J1757$-$2745, and J1804$-$2858. For comparison, we also plot $\dot{P}_{\rm obs}/P$ of known pulsars ($grey$ dots), and marked them with ${\it black}$ dots and include their names in {\it black} if they are comparable (less than a factor of four) to the model-dependent maximum acceleration. Note that if the pulsar is located in the fourth quadrant of $l$, then we plotted $360^\circ-l$. For comparison, we also include the maximum LOS acceleration curves ({\it grey} curves) obtained from the Galactic model given in \citet{mcm17}.
}
\label{results}
\end{figure}

We estimate the maximum possible LOS acceleration curves $a_{\rm max}/c$ due to the Galactic potential as a function of Galactic longitude for a given Galactic latitude (see Fig.~\ref{results}). We plot the absolute values of $|\dot{P}_{\rm obs}/P|$ of PSRs~J1748$-$3009, J1753$-$2819, J1757$-$2745, and J1804$-$2858 in the figure. Due to the axially symmetric mass distribution and the acceleration, we map the pulsars that are located in the fourth quadrant of the Galaxy in the first quadrant and include their $\dot{P}_{\rm obs}/P$ in the  figure. It is evident that, as shown in Fig.~\ref{results}, all our six PSRs~J1746$-$2758, J1748$-$3009, J1753$-$2819, J1757$-$2745, J1801$-3210$, and J1804$-$2858 have much lower $|\dot{P}_{\rm obs}/P|$ than the maximum Galactic accelerations along their LOSs obtained from our main Galactic model \citep{sof13}. Therefore, we cannot simply neglect the Galactic contribution in their observed $\dot{P}$. We investigate the LOS accelerations of these pulsars in detail in \S~\ref{pdot}, leading to limits on their intrinsic $\dot P$ values. Fig.~\ref{results} also shows that most of the known pulsars that are located within the same field of view have much larger $\dot{P}_{\rm obs}/P$ compared to their Galactic maximum accelerations. If their Shklovskii terms are small, we can simply assume that the observed $\dot P$ measurements of these pulsars are approximately equal to their intrinsic values. In addition, PSRs~J1723$-$2837, J1727$-$2946, J1751$-$2857, and J1804$-2717$ have observed $\dot{P}_{\rm obs}/P$ that are comparable to the maximum Galactic accelerations along their LOSs (see {\it black} dots in Fig.~\ref{results}). We note that the DM measurements of these pulsars are relatively small \citep[between $19-60$~cm$^{-3}$~pc;][]{hlk+04,cls+13,lem+15,dcl+16} compared to those of our pulsars in the analysis (see Table~\ref{psr_info}). The DM-derived distances (using YMW16 model) to these pulsars are in the range of $0.72-1.9$~kpc (note that the NE2001 model derived distances are also in a similar range). This indicates that they are located close to the Sun and thus, the Galactic contributions in their observed $\dot{P}$ values are insignificant, which means that the observed $\dot P$ values are approximately equal to their intrinsic values.

\subsection{Constraining the intrinsic period derivatives of pulsars}
\label{pdot}

\begin{figure}
\includegraphics[width=8cm]{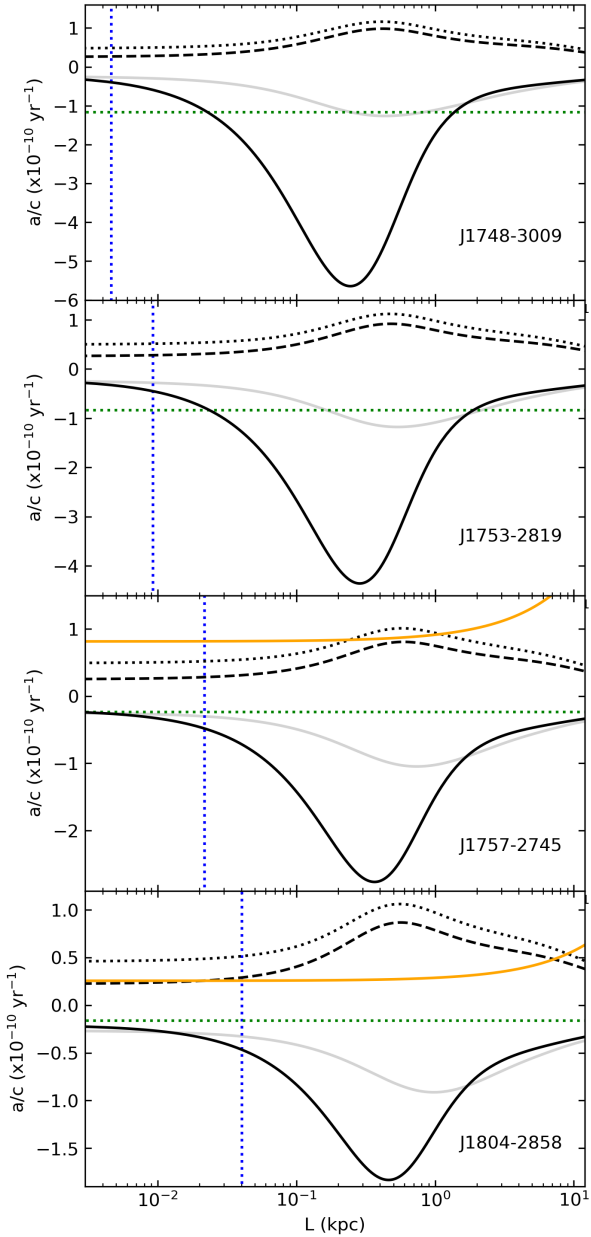}
\centering
\caption{
The Galactic LOS acceleration values of PSRs~J1748$-$3009, J1753$-$2819, J1757$-$2745, and J1804$-$2858 as a function of distance $L$ measured from the sky plane (see Fig.~\ref{diagram}) along their LOSs based on \citet{sof13} ({\it black} solid) and \citet{mcm17} ({\it grey} solid). The {\it green} dotted line represents the observed $\dot{P}/P$ based on the timing measurements. The {\it blue} dotted line represents the distance measured from the sky plane of the pulsar to the plane which passes through the GC normal to the line joining the observer and the GC -- i.e. $r_\perp \tan(\theta)$ (see Fig.~\ref{diagram}). The {\it black} dashed and dotted lines represent the expected Shklovskii contribution for an assumed 100~km/s and 320~km/s space velocities, respectively. The estimated Shklovskii contribution based on the measured proper motion limits of PSRs~J1757$-$2745 and J1804$-$2858 are over-plotted separately ({\it orange} line).
}
\label{acc_l}
\end{figure}

We know that $\dot{P}_{\rm int}$ and thus, $\dot{P}_{\rm int}/P$, of a non-accreting pulsar is positive. As described before, $a_\mu/c$ is a positive contribution and thus, the measured negative $\dot P_{\rm obs}$ measurements of PSRs~J1748$-$3009, J1753$-$2819, J1757$-$2745, and J1804$-$2858 can only be explained with a negative $a/c$ Galactic contribution (see Eq.~\ref{acc_eq}). Using a Galactic model as shown in \S~\ref{galactic_model}, we can compute the minimum possible Galactic acceleration $(a_{\rm min}/c)$ along a given LOS. Therefore, we can write the upper limit of the intrinsic period derivative for these four pulsars to be
\begin{equation}
\label{upper}
\left(\frac{\dot{P}_{\rm int}}{P}\right)_{\rm upper} = \frac{\dot{P}_{\rm obs}}{P} - \frac{a_{\rm min}}{c} - \frac{a_\mu}{c}
\end{equation}
using Eq.~\ref{acc_eq}. 
We first estimate the Galactic acceleration curves of these pulsars along their LOSs and plot in Fig.~\ref{acc_l}. We then estimate $a_\mu/c$ assuming the pulsar moves with a space velocity of either 100~km/s (black {\it dashed} line) or 320~km/s (black {\it dotted} line) and over-plot in the same figure. For clarity, the x-axis in Fig.~\ref{acc_l} represents the distance $L$ measured from the sky-plane (see Fig.~\ref{diagram} for definitions). Appendix~\ref{velocity} describes the calculation of $a_\mu/c$ in detail. We then estimate the model-dependent upper limit on the intrinsic $\dot {P}$ of PSRs~J1748$-$3009, J1753$-$2819, J1757$-$2745, and J1804$-$2858 using Eq.~\ref{upper} with the Galactic accelerations shown in Fig.~\ref{acc_l}. The estimated limits are given in Table~\ref{pdot_int}. PSRs J1757$-$2745 and J1804$-$2858 have measured values on their proper motions (see Table~\ref{sol_all_2}). For comparison, we estimate the Shklovskii contribution based on these measured proper motions in Right Ascension and over-plot in Fig.~\ref{acc_l} (see the {\it orange} curves).

\begin{table}
\begin{center}
\caption{
The derived parameters of pulsars in the study. The minimum Galactic acceleration $a_{\rm min}/c$ is obtained from the Galactic model given in \citet{sof13}. The model-dependent upper limit on the intrinsic $\dot P$ is derived using Eq.~\ref{upper} (see \S~\ref{pdot}). The last column represents the model-dependent distance to the pulsar determined through dynamics as described in \S~\ref{location}. Note that for PSRs~J1757$-$2745, J1801$-$3210, and J1804$-$2858, we quote the derived distances obtained using their timing measured proper motion limits.
 }
\label{pdot_int}
\begin{tabular}{lccc}
\hline
\multicolumn{1}{c}{PSR} &
\multicolumn{1}{c}{$a_{\rm min}/c$} &
\multicolumn{1}{c}{$\dot{P}_{\rm int}$} & 
\multicolumn{1}{c}{$D$} \\
\multicolumn{1}{c}{ } &
\multicolumn{1}{c}{($10^{-10}$ yr$^{-1}$)} &
\multicolumn{1}{c}{($10^{-19}$ s/s)} &
\multicolumn{1}{c}{(kpc)} \\
\hline
\hline
J1746$-$2758 & $-6.70$ & $<$148 & $<$7.50 or $>$8.16 \\
J1748$-$3009 & $-5.64$ & $<$1.38 & $8.24-8.99$ \\
J1753$-$2819 & $-4.36$ & $<$2.08 & $8.24-9.08$ \\
J1757$-$2745 & $-2.77$ & $<$1.42 & $8.27-9.40$ \\
J1801$-$3210 & $-1.48$ & $<$0.35 & $8.20-15.50$   \\
J1804$-$2858 & $-1.83$ & $<$0.08 & $8.32-9.40$ \\
\hline
\end{tabular}
\end{center}
\end{table}

In addition to the observed $\dot{P}$, the timing solution of PSR~J1748$-$3009 places a limit on its orbital period derivative $\dot{P}_b$ (see Table~\ref{sol_all_1}). The timing parameters indicate that the pulsar is in a compact binary system with a period of 2.9-d and possibly in orbit with a white draft (the mass of the companion is $0.08-0.20$~M$_\odot$ with a median of 0.1~M$_\odot$). Therefore, we can assume that the observed $\dot{P}_b$ includes contributions from both dynamics and the emission of gravitational waves \citep[see][]{tw89,sttw02,ksm+06}. Thus, the observed $\dot{P}_b$ provides important information to estimate the intrinsic period derivative of the pulsar independently. This approach has been used previously in several studies \citep[e.g.,][]{mnf+16,frk+17,prf+17}. Similar to Eq.~\ref{acc_eq}, we can express the pulsar LOS acceleration in terms of observed $\dot{P}_b$ as follows:
\begin{equation}
\label{pbdot}
\frac{\dot{P}_{\rm b}}{P_{\rm b}} = \left(\frac{\dot{P}_{\rm b}}{P_{\rm b}}\right)_{\rm GR} + \frac{a}{c} + \frac{a_{\rm \mu}}{c},
\end{equation}  
where $(\dot{P}_{\rm b}/P_{\rm b})_{\rm GR}$ is the expected contribution due to gravitational wave emission. Assuming General Relativity, the expected $(\dot{P}_b)_{\rm GR}$ is $-4.3\times10^{-16}$~s/s \citep[using Eq.~8.52 in][assuming a pulsar mass of 1.4~M$_\odot$ and $\cos(i)=0.5$, resulting in a companion mass of 0.1~M$_\odot$]{lk05}, which is significantly smaller than the observed $\dot{P}_b$ limit of $-2.6(10)\times10^{-11}$~s/s. This indicates that the origin of the observed $\dot{P}_b$ is most likely induced by the dynamical acceleration component of the pulsar along the LOS, similar to the observed $\dot P$ measurement discussed earlier. Therefore, we can simply assume that $(a/c + a_\mu/c)\approx \dot{P}_{b}/P_b$ from Eq.~\ref{pbdot} and then substitute that in to Eq.~\ref{acc_eq} to deduce $\dot{P}_{\rm int}$ independently, and it is calculated to be $(9.7\pm8.0)\times10^{-19}$~s/s, where the uncertainty is 2$\sigma$. 
Note that this measurement is greater than the upper limit constrained above using the Galactic mass model (see Table~\ref{pdot_int}). 
The reason for this discrepancy could be that we only have a limit on $\dot{P}_b$ through timing, so that the above estimated $\dot{P}_{\rm int}$ from Eq.~\ref{pbdot} is not well constrained.

\begin{figure}
\includegraphics[width=8cm]{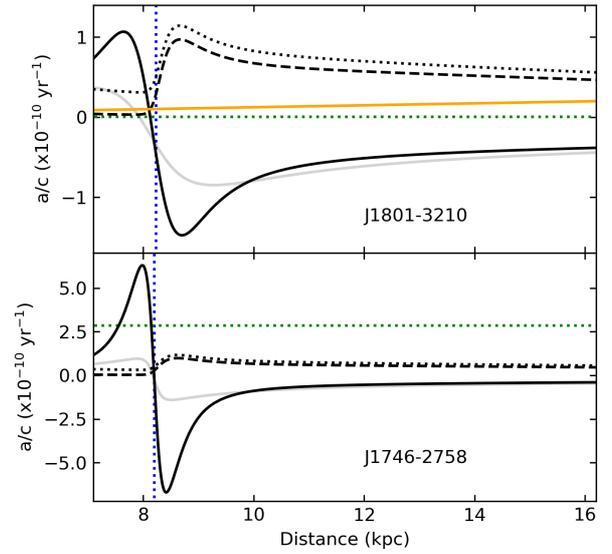}
\centering
\caption{
Same as Fig.~\ref{acc_l}, but for PSRs~J1801$-$3210 and J1746$-$2752. Note that the x-axis is the distance measured from the Sun along the pulsar LOS.
}
\label{acc_l_1801_1746}
\end{figure}

As given in Table~\ref{sol_all_2}, the timing model of PSR~J1801$-$3210 shows a positive, but extremely small $\dot{P}$.
Given that PSR~J1801$-$3210 is located towards the GC (with an angular separation of $4\fdg7$ -- see Table~\ref{summary}) and has a large DM, it is probable that the measured $\dot{P}$ is dominated by the acceleration of the pulsar due to the Galactic potential.
If $\dot{P}_{\rm obs} < \dot{P}_{\rm int}$, the contribution $a/c$ must be negative (see Eq.~\ref{acc_eq}). We plot the acceleration components given in Eq.~\ref{acc_eq} for this pulsar along its LOS in Fig.~\ref{acc_l_1801_1746}. Following the same method as above, we can then estimate the model-dependent upper limit on its intrinsic $\dot P$ using Eq.~\ref{upper}, leading to a value of $3.5\times10^{-20}$~s/s (see Table~\ref{summary}). This limit is broadly consistent with the measured $\dot P$ values of unaccelerated millisecond pulsars with a similar period and we suspect that it is closer to the intrinsic $\dot P$ value of the pulsar.

PSR~J1746$-$2758 has a positive $\dot{P}$ measurement of $4.42(5)\times10^{-18}$~s/s, but this value is slightly smaller than that of the majority of other known pulsars that have a similar period (\S~\ref{J1746}). Given that it is located towards the GC, with an angular separation of $0\fdg96$, and also has a large DM, it is possible that the observed $\dot P$ of PSR~J1746$-$2752 is also influenced by its LOS acceleration. Therefore, we estimate the acceleration curves of this pulsar following the same procedure as described above (see Fig.~\ref{acc_l_1801_1746}). Using Eq.~\ref{upper}, we then estimate the upper limit on its intrinsic $\dot P$ to be $1.48\times10^{-17}$~s/s, which is approximately a factor of 3 larger than the measured value.

\begin{figure}
\includegraphics[width=8.5cm]{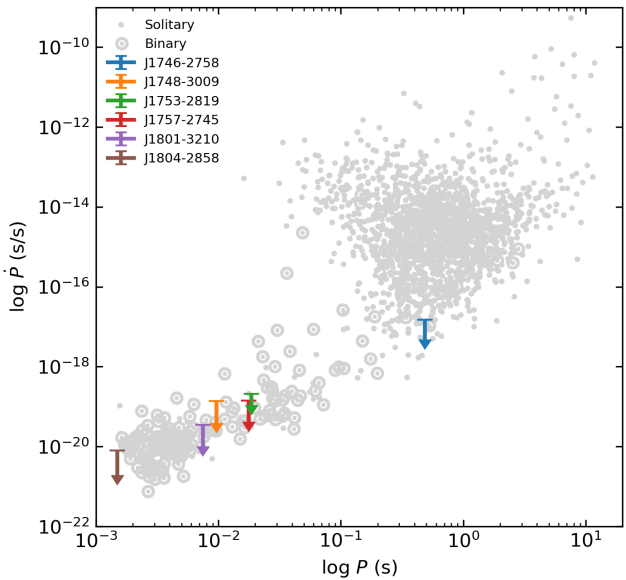}
\centering
\caption[]{
The $P$ and $\dot{P}$ diagram generated using all the known pulsars in the catalogue\footnotemark. The derived model-dependent intrinsic period derivative limits of the pulsars in this analysis as given in Table~\ref{pdot_int}. We note that these limits are broadly consistent with the known pulsar population.
}
\label{ppdot}
\end{figure}

We finally plot the dynamically-derived model-dependent intrinsic $\dot P$ limits of all our pulsars with their measured $P$ values together with those of other known pulsars in Fig.~\ref{ppdot}. It is clear that the dynamically-derived $\dot P$ limits of our pulsars are broadyly consistent with those of other pulsars. Using these upper limits on $\dot P$ values, we estimate and present the lower limits on characteristic ages and upper limits on surface magnetic fields in Table~\ref{sol_all_1} and \ref{sol_all_2} \citep[using Eq.~3.12 and 3.15 given in][]{lk05}.

\footnotetext{\url{http://www.atnf.csiro.au/research/pulsar/psrcat/}}

\subsection{Dynamical constraints on pulsar distances}
\label{location}

\begin{figure}
\includegraphics[width=8cm]{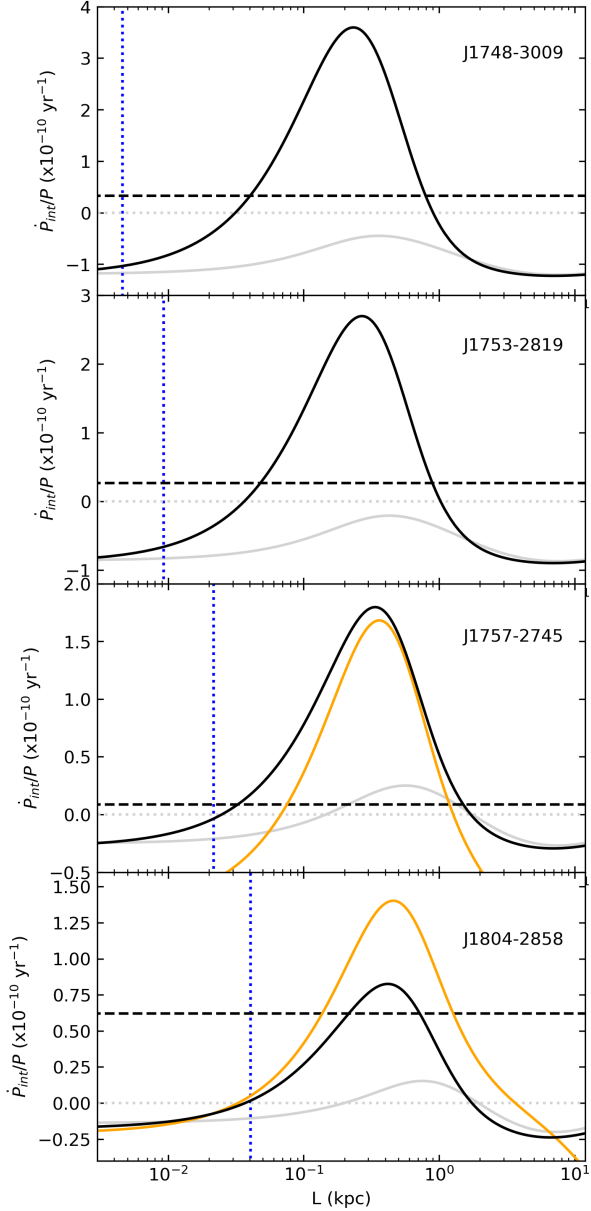}
\centering
\caption{
The model-estimated values of $\dot{P}_{\rm int}/P$ obtained using  Eq.~\ref{acc_eq} based on dynamic acceleration and Shklovskii contributions given in Fig.~\ref{acc_l} for PSRs~J1748$-$3009, J1753$-$2819, J1757$-$2745, and J1804$-$2858 ({\it black} solid -- \citet{sof13}; {\it grey} solid -- \citet{mcm17}). Note that a pulsar space velocity of 100~km/s is assumed in the estimation of $\dot{P}_{\rm int}/P$. The {\it blue} dotted line represents the distance measured from the sky plane to the plane which passes through the GC normal to the line joining the observer and the GC (see Fig.~\ref{diagram}). The dashed {\it black} horizontal line represents the lowest measured $\dot{P}/P$ value determined from the known population of pulsars around the spin period of the particular pulsar (see the text), and the distance range is estimated when $\dot{P}_{\rm int}/P$ is greater than this value. The estimated distances are given in Table~\ref{summary} based on the \citet{sof13} model. 
Note that L represents the distance measured from the sky plane along the LOS. To get the pulsar distance, the distance to the sky plane from the Sun along the LOS (i.e. $R_{\odot}\cos(\theta)$, where $R_{\odot}=8.2$~kpc and $\theta$ is the angular separation of the pulsar -- see Table~\ref{summary}) should be added to L (see Fig.~\ref{diagram}).
The {\it orange} lines present the derived $\dot{P}_{\rm int}/P$ curve using the Shklovskii contribution estimated from the measured proper motion limits of PSRs~J1757$-$2745 and J1804$-$2858. 
}
\label{pdot_int_l}
\end{figure}

\begin{figure}
\includegraphics[width=8cm]{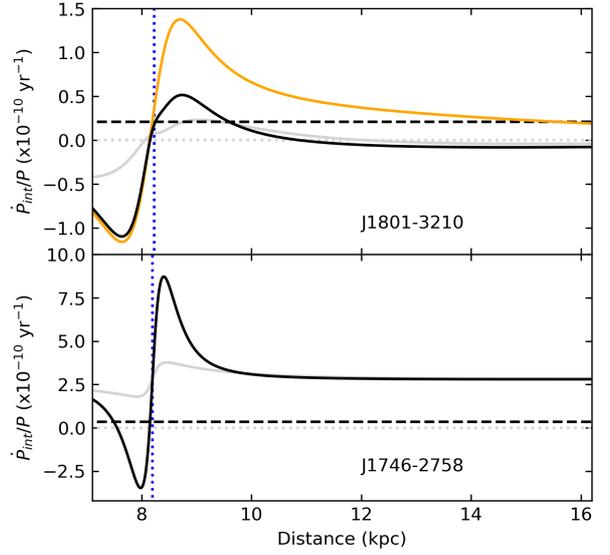}
\centering
\caption{
Same as Fig.~\ref{pdot_int_l}, but for PSRs~J1801$-$3210 and J1746$-$2758. Note that the x-axis represents the distance measured from the Sun along the pulsar LOS.
}
\label{pdot_1801_1746}
\end{figure}

Using the LOS acceleration values shown in Figures~\ref{acc_l} and \ref{acc_l_1801_1746}, we can simply estimate the intrinsic period derivatives $\dot{P}_{\rm int}$ of PSRs~J1746$-$2758, J1748$-$3009, J1753$-$2819, J1757$-$2745, J1801$-$3210, and J1804$-$2858 as a function of distance using Eq.~\ref{acc_eq} (see Fig.~\ref{pdot_int_l} and \ref{pdot_1801_1746}). By definition we know that $\dot{P}_{\rm int}$ is positive and thus, these curves can be used to constrain the model-dependent distances to these pulsars. Using the known pulsar population, we can impose possible lower limits on their $\dot{P}_{\rm int}$ values. To do that, we select known pulsars within a 50 per cent window around the period of a given pulsar in our sample, and then obtain the lowest observed $\dot P$ in that population. This lowest $\dot P$ is used as the minimum possible $\dot{P}_{\rm int}$ for that given pulsar and it is plotted in Fig.~\ref{pdot_int_l} and \ref{pdot_1801_1746} with horizontal {\it dashed} lines. Assuming that $\dot{P}_{\rm int}$ should be greater than this minimum limit, we constrain the distances to be $(8.24-8.99)$, $(8.24-9.08)$, $(8.22-9.65)$, $(8.25-9.64)$, and $(8.40-8.90)$~kpc for PSRs~J1748$-$3009, J1753$-$2819, J1757$-$2745, J1801$-$3210, and J1804$-2858$, respectively (see Table~\ref{summary}). We notice that the lower limits of these distances are beyond the distances to the plane which passes through the GC perpendicular to the line joining the Sun and the GC along their LOSs, placing all these five pulsars on the far-side of the Galaxy. The distance to PSR~J1746$-$2758 is not well constrained (see Fig.~\ref{pdot_1801_1746}), although, it is likely to be located at a distance of either $<$7.5~kpc or $>$8.16~kpc. 
In all the above distance estimates, we assume \cite{sof13} as our main Galactic model and the peculiar velocity of the pulsar to be 100~km/s when calculating the Shklovskii contribution. If the velocity is larger than this assumed value, then $\dot{P}_{\rm int}$ decreases (i.e. the intrinsic curve shifts downward along y-axis) and thus, constrains the distance to a smaller range.
Moreover, using the timing-measured proper motion limits of PSRs~J1757$-$2745, J1801$-$3210, and J1804$-$2825, we estimate the Shklovskii contribution and then derive $\dot{P}_{\rm int}$ as before (see {\it orange} curves in Fig.~\ref{pdot_int_l} and \ref{pdot_1801_1746}), leading to more conservative pulsar distances of $(8.27-9.40)$, $(8.20-15.50)$, and $(8.32-9.40)$~kpc, respectively.

Note that, as we mentioned in \S~\ref{galactic_model}, the \citet{mcm17} model mainly focuses beyond the bulge regions and thus, it cannot be used accurately to constrain the distances to these pulsars. This can be seen in Fig.~\ref{pdot_int_l} and \ref{pdot_1801_1746} as the model-derived $\dot{P}_{\rm int}$ values ({\it grey} solid curves) are potentially smaller than the minimum possible $\dot{P}_{\rm int}$ values (dashed {\it black} horizontal lines).

\subsection{Comparison of dynamic-derived distances with DM-derived distances}
\label{comparison}

\begin{figure}
\includegraphics[width=8.3cm]{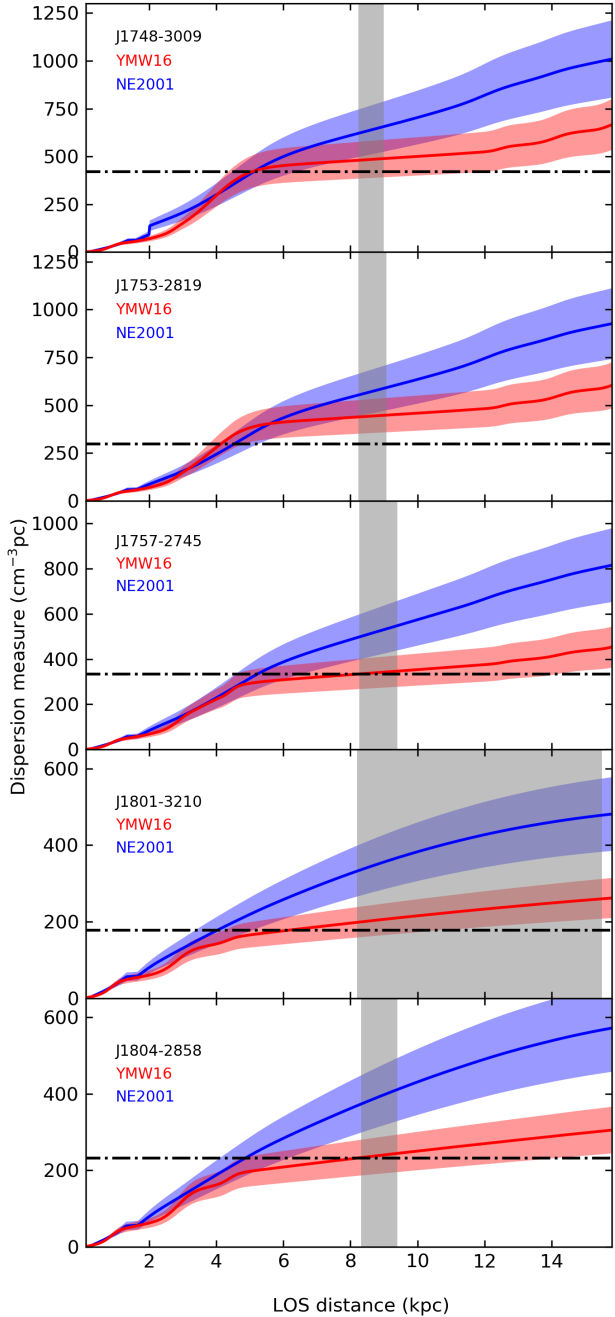}
\centering
\caption{
The DM values obtained from YMW16 ({\it red}) and NE2001 ({\it blue}) electron density models for our pulsars as a function of distance along their LOSs. The {\it shaded blue} and {\it red} regions represent the typical 20 per cent uncertainty of these models. The horizontal {\it dot-dashed} line represents the DM value of the pulsar (see Table~\ref{psr_info}). The {\it grey} {\it shaded} region represents the constrained model-dependent distances to these pulsars through our dynamic analysis  (see \S~\ref{location} and Table~\ref{pdot_int}).     
}
\label{dm_dist}
\end{figure}

The thermal free electron density model NE2001\footnote{\url{https://www.nrl.navy.mil/rsd/RORF/ne2001}} \citep{cl02} is widely used to estimate the distances to pulsars based on their measured DM obtained through timing. More recently, \citet{ymw17} (YMW16\footnote{\url{http://119.78.162.254/dmodel/index.php}}) re-evaluated the electron density model by using more recent independent pulsar distance observations. We compare our dynamically-estimated distances for PSRs~J1748$-$3009, J1753$-$2819, J1757$-$2745, J1801$-$3210, and J1804$-$2825 given in \S~\ref{location} with their DM-derived distances obtained from these electron density models. Note that we do not use PSR~J1746$-$2758 in this comparison since its distance is not well constrained. We first use the YMW16 and NE2001 models separately to estimate the DM along the LOS of these pulsars (see Fig.~\ref{dm_dist}). Note that we take account of a typical 20 per cent error on these DM values and indicate it with the shaded region in the figure. Based on the measured DM values, we find that the DM-derived distances obtained from the YMW16 model for PSRs~J1748$-$3009, J1757$-$2745, J1801$-$3210, and J1804$-$2825 are consistent with our dynamics-derived distances within their uncertainties. For PSR~J1753$-$2819, the DM-derived distance from YMW16 is small compared to its dynamics-derived distance. This indicates that the electron density predicted from YMW16 along the LOS of PSR~J1753$-$2819 is overestimated, resulting in underestimated distance.
As shown in the figure, the DM-derived distances obtained from NE2001 model for all of these pulsars are underestimated compared to their dynamics-estimated distances. 
By comparing the two models, it is seen that in general the distances estimated by the NE2001 model are underestimated \citep[as reported before, see, e.g.,][]{kbm+03} compared to those estimated by the YMW16 model at a given DM along the LOSs of these pulsars.

\begin{figure}
\includegraphics[width=8.4cm]{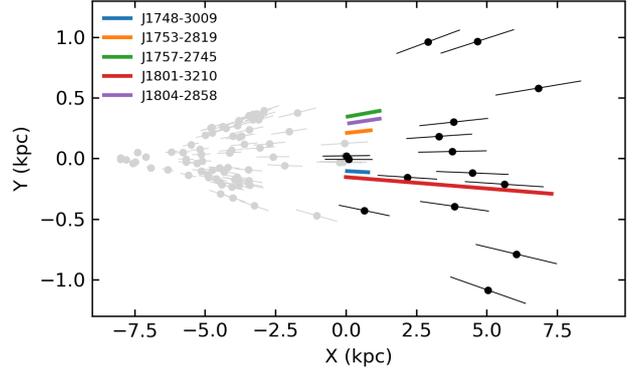}
\centering
\caption{
The observed pulsars located within an angular radius of $5^\circ$ in the direction of the GC. A projection onto the Galactic plane is shown in cartesian coordinates. Note that the GC is at $(0,0)$ and the Sun is at $(-8.2,0.0)$~kpc. The distances are obtained from the YMW16 model for pulsars that are located on the near-side ($\it grey~dots$) and the far-side ($\it black~dots$), separately. The typical 20 per cent uncertainties of the distances are shown. The distances obtained from dynamics for PSRs~J1748$-$3009, J1753$-$2819, J1757$-$2745, J1801$-$3210, and J1804$-$2858 in this study are over-plotted with different colours.
}
\label{psr_galaxy}
\end{figure}

We consider all known pulsars that are located within an angular radius of $5^\circ$ from the GC (i.e. $\theta < 5^\circ$) and plot their projected distances in the Galactic plane (see Fig.~\ref{psr_galaxy}). Note that, we use DM-derived distances for these pulsars using the YMW16 model and assume a typical 20\% uncertainty. We also plot our dynamically-derived distances for PSRs~J1748$-$3009, J1753$-$2819, J1757$-$2745,  J1801$-$3210, and J1804$-$2858 in the same figure. As can be seen, our distance estimates have much smaller uncertainties compared to the DM-derived distances, and these five pulsars are located on the far-side of the Galaxy.

\section{Discussion and Conclusion}
\label{concl}

We report on the timing of recently discovered PSR~J1753$-$2819, a 18.6~ms pulsar in a 9.3-hr binary orbit. We included this pulsar along with PSRs~J1746$-$2758, J1748$-$3009, J1757$-$2745, J1801$-$3210, and J1804$-$2858 using LT and PKS observations to understand their $\dot P$ measurements. The improved timing baselines indicate significant negative $\dot P$ measurements for PSRs J1748$-$3009, J1753$-$2819, J1757$-$2745, and J1804$-$2858 (see Table~\ref{sol_all_1} and \ref{sol_all_2}). This is the first non-globular cluster pulsar sample that has observed negative $\dot P$ measurements in their timing solutions, and they all lie within a few degrees of the GC. In addition, the timing of PSR~J1801$-$3210 confirmed that it has an extremely low positive $\dot P$ measurement, which provides the lowest measured value for a pulsar to date.

Since pulsars are powered by the loss of rotational kinetic energy, they spin-down gradually with a positive intrinsic $\dot P$.
Therefore, the negative $\dot P$ measurements of PSRs~J1748$-$3009, J1753$-$2819, J1757$-$2745, J1804$-$2858 and the extremely small $\dot P$ measurement of PSR~J1801$-$3210 are not intrinsic, but induced by dynamics. We modelled the observed $\dot P$ values of these pulsars as being due to dynamical acceleration in the Galactic potential along their LOSs. The analysis constrained the model-dependent upper limits on the intrinsic $\dot P$ of these pulsars and they are broadly consistent with those of known pulsars (see Fig.~\ref{ppdot} and Table~\ref{pdot_int}). 
Using these model-dependent intrinsic $\dot P$ values, we then constrained possible pulsar distances, indicating that PSRs~J1748$-$3009, J1753$-$2819, J1757$-$2745, J1801$-$3210, and J1804$-$2858 are located on the far-side of the Galaxy (see Fig.~\ref{psr_galaxy} and Table~\ref{pdot_int}). This is the first time that dynamics have been used to measure the distances to pulsars that are located in the far-side of the Galaxy.

The timing solutions of three pulsars in our sample measured limits on their proper motions (see Table~\ref{sol_all_2}). Assuming the Galactic model \citep{sof13} is correct and $\dot{P}_{\rm int}>0$, we can determine the maximum possible total proper motion for PSRs~J1757$-$2745, J1801$-$3210, and J1804$-$2858 and they are estimated to be 19.2, 12.6, and 13~mas/yr (using Eq.~\ref{acc_eq}). These maximum possible total proper motions are consistent with our timing-measured limits given in Table~\ref{sol_all_2}. If the high-significance timing-measured proper motions in the future for these pulsars are greater than the above mentioned values, we can validate the Galactic model and also include any required additional features in the model to explain the observed pulsar motions.

As shown in Fig.~\ref{dm_dist}, our dynamics-derived distances and the DM-derived distances from the YMW16 electron density model for these pulsars are consistent with each other in general within their uncertainties. However, the NE2001 model-derived distances are inconsistent with the dynamics-derived distance. This could be due to difficulties in modelling the electron densities in the inner regions of the Galaxy with high stellar densities and large gas and dust contents towards the centre. Therefore, the independent distance measurements obtained to the pulsars in this study can  be used to further improve electron density models.

PSRs~J1748$-$3009, J1753$-$2819, and J1801$-$3210 are in binary systems and the timing results indicate that their companions are likely to be white dwarfs. Using the measured $\dot{P}_b$ limit of PSR~J1748$-$3009, we derived an independent limit on its intrinsic $\dot P$. Following a similar method, we will be able to measure $\dot{P}_b$ values of these binary pulsars in the future with more observations and thus, derive their intrinsic $\dot P$ values independently with better accuracy \citep[see][]{stw+11}. These estimates then can be used to improve their distance measurements.

Most of the Galactic models assume that the mass distribution in the Galaxy is either axially or spherically symmetric for simplicity, in particular within the bulge region. Having more pulsars in the Galactic centre region, we can use their observations to derive dynamics independently to study the Galactic potential, and then update and validate Galactic models including clumpiness in the mass distribution. We can also obtain independent distances to these GC pulsars in the future through parallax measurements using interferometers such as VLBI and then use them in deriving the Galactic potential. The next generation telescopes, such as the Square Kilometre Array, are useful with their outstanding capabilities to discover more pulsars in the GC region and to allow a comprehensive study to constrain the Galactic potential in the central region using pulsar timing in the future.

\section*{Acknowledgments}
Pulsar research at the Jodrell Bank Centre for Astrophysics and the observations using the Lovell Telescope are supported by a consolidated grant from the STFC in the UK. The Parkes radio telescope is part of the Australia Telescope, which is funded by the Commonwealth of Australia for operation as a National Facility managed by the Commonwealth Scientific and Industrial Research Organisation (CSIRO). D.R.L. acknowledges support from NSF RII Track I award number OIA$-$1458952 and is part of the NANOGrav Physics Frontiers Center which is supported by NSF award 1430284. R.S. acknowledges support from the Australian Research Council grant FL150100148. MB acknowledges ARC Grant CE170100004 (OzGrav).

\bibliography{psrrefs,modrefs,journals}
\bibliographystyle{mnras}

\appendix
\section{Transverse velocity of the pulsar and the Shklovskii contribution}
\label{velocity}

We denote the peculiar velocity of the Sun as $\vec{v}_{_\odot,{\rm pec}} = [U, V, W]$, where the unit vectors of the components are in the direction from the Sun towards the GC, in the direction of the Galactic rotation, and in the direction perpendicular to the Galactic plane, respectively. Then the total velocity of the Sun 
\begin{equation}
\vec{v}_\odot = [U, V+v_R(R_\odot), W],
\end{equation}
where $v_R(R_\odot)$ is the velocity of the local standard of rest for the Sun, or the circular Galactic rotational velocity at the radius of Sun $R_\odot$. For our calculation, we assume $[U, V, W] = [11.1, 12.24, 7.25]$~km/s \citep{sbd10} and obtain the Galactic rotation $v_R$ from Galactic mass distribution models given in \S~\ref{galactic_model}. The main Galactic model we use in this study  \citep{sof17} results in $v_R(R_\odot) = 238$~km/s, which is the value they adopted in their model from \citet{hna+12}. Then the peculiar velocity, or the 3-dimensional space velocity, of the pulsar is defined as $\vec{v}_{\rm p, pec} = [U_p, V_p, W_p]$ in the same coordinate system as before. The total velocity of the pulsar including its local standard of rest,
\begin{equation}
\vec{v}_{\rm psr} = [U_p + v_R(R_p)\sin\theta', V_p+v_R(R_p)\cos\theta', W_p],
\end{equation}
where $v_R(R_p)$ is the Galactic rotational velocity of the pulsar at its projected radial distance $R_p$ on to the Galactic plane and $\theta'$ is the angle subtended at the GC between the position of the Sun and the projected position of the pulsar on the Galactic plane \citep[see Fig.~B.1 in][for a schematic diagram]{vic17}. The angle $\theta'$ can be expressed as
\begin{equation}
\tan (\theta' + l) = \frac{R_\odot \sin l}{R_\odot  \cos l - d \cos b}.
\end{equation} 

Using the above coordinate system, the relative velocity of the pulsar with respect to the Sun
\begin{equation}
\begin{split}
\vec{v}_{\rm psr} - \vec{v}_\odot = ~& [U_p + v_R(R_p)\sin\theta' - U, \\ & V_p+v_R(R_p)\cos\theta' - V-v_R(R_\odot), W_p - W].
\end{split}
\end{equation}
To express the relative velocity in the Galactic coordinate system $(l, b)$, we write the unit vectors $(\hat{l}, \hat{b})$ in above coordinate system as  
\begin{equation}
\begin{split}
\hat{l} &= [-\sin l, \cos l, 0] \\
\hat{b} &= [-\cos l \sin b, -\sin l \sin b, \cos b].
\end{split}
\end{equation}
We then write the relative velocity of the pulsar for its given $(l, b)$ with respect to the Sun in Galactic coordinates as
\begin{equation}
\label{vl}
\begin{split}
v_l = ~& \hat{l}\cdot(\vec{v}_{\rm psr} - \vec{v}_\odot ) \\
 = ~& -(U_p + v_R(R_p)\sin\theta' - U)\sin l \\
 &+(V_p+v_R(R_p)\cos\theta' - V-v_R(R_\odot))\cos l 
\end{split}
\end{equation}
and
\begin{equation}
\label{vb}
\begin{split}
v_b = ~& \hat{b}\cdot(\vec{v}_{\rm psr} - \vec{v}_\odot ) \\
 = ~& -(U_p + v_R(R_p)\sin\theta' - U)\cos l \sin b \\
 &-(V_p+v_R(R_p)\cos\theta' - V-v_R(R_\odot))\sin l \sin b \\
 &+(W_p - W)\cos b,
\end{split}
\end{equation}
respectively. We finally calculate the transverse velocity of the pulsar to be $v_T = \sqrt{(v_l^2 + v_b^2)}$.
The derivation of the above expressions are given in \citet{vic17} in detail.

In \S~\ref{dynamics}, for a given 3-dimensional pulsar space velocity of $|\vec{v}_{\rm p,pec}|$, we randomly calculate $[U_p, V_p, W_p]$ (where $|\vec{v}_{\rm p,pec}| = \sqrt{U_p^2 + V_p^2 + W_p^2}$), leading to a transverse velocity of $v_T$ using Eq.~\ref{vl} and \ref{vb}. We perform 1000 trials for a given $|\vec{v}_{\rm p,pec}|$ and then obtain the average $v_T$, and then calculate the Shklovskii effect $a_\mu /c = v_T^2/cd$. We note that $v_T$ is a function of $R_p$ and thus, it varies along the LOS of the pulsar for a given $\vec{v}_{\rm p,pec}$, and then $a_\mu /c$. For instance, this variation in $a_\mu /c$ can be seen in Fig.~\ref{acc_l}.

\end{document}